\documentclass[aps,rmp,reprint,amsmath,amssymb,floatfix,longbibliography,amsart]{revtex4-1}
\usepackage[pdftex]{color,graphicx}
\usepackage{bm}
\usepackage{natbib}
\usepackage{verbatim}
\usepackage{booktabs,multirow}
\usepackage[table]{xcolor}
\usepackage{mathrsfs}
\usepackage{amssymb}
\usepackage{exscale}
\providecommand{\keywords}[1]{\textbf{\textit{Keywords ---}} #1}
\newcommand{\beq}{\begin{equation}}
\newcommand{\eeq}{\end{equation}}

\definecolor{applegreen}{rgb}{0.55, 0.71, 0.0}
\definecolor{amethyst}{rgb}{0.6, 0.4, 0.8}
\definecolor{tangelo}{rgb}{0.98, 0.3, 0.0}
\newcommand{\sk}[1]{\textcolor{black}{#1}}
\newcommand{\qs}[1]{{\color{black}#1}}

\begin{document}

\title{Heavy-electron quantum criticality and single-particle spectroscopy}

\author{Stefan Kirchner}\email{stefan.kirchner@correlated-matter.com}
\affiliation{Zhejiang Institute of Modern Physics, Zhejiang University, Hangzhou,  Zhejiang 310058, China\\ and Zhejiang Province Key Laboratory of Quantum Technology and Device, Zhejiang University, Hangzhou 310027, China}

\author{Silke Paschen}
\affiliation{Institute of Solid State Physics, Vienna University of Technology, Wiedner 
	Hauptstrasse
	8-10, 1040 Vienna, Austria}

\author{Qiuyun Chen}
\affiliation{Science and Technology on Surface Physics and Chemistry Laboratory, Mianyang 621908, China}

\author{Steffen Wirth}
\affiliation{Max Planck Institute for Chemical Physics of Solids, 01187 Dresden, Germany}

\author{Donglai Feng}
\affiliation{State Key Laboratory of Surface Physics, and Department of Physics, Fudan University, Shanghai 200433, China\\
and Hefei National Laboratory for Physical Science at Microscale, CAS Center for Excellence in Quantum Information and Quantum Physics,
and Department of Physics, University of Science and Technology of China, Hefei 230026, China}

\author{Joe D. Thompson}
\affiliation{Los Alamos National laboratory, Los Alamos, New Mexico 87545, USA}

\author{Qimiao Si}
\affiliation{Department of Physics and Astronomy
	and
	Center for Quantum Materials, Rice University, Houston, Texas 77005, USA}

\date{\today{}}
\begin{abstract} 
Angle-resolved photoemission spectroscopy (ARPES) and scanning tunneling microscopy (STM) 
have become indispensable tools in the study of correlated quantum materials. 
Both probe  complementary aspects of the single-particle excitation spectrum.  
Taken together, ARPES and STM  have the potential to explore properties of the electronic Green's function, 
a central object of many-body theory.
This review explicates this potential with a focus on heavy-electron quantum criticality, 
especially
the role of Kondo
destruction.
A discussion on  how to 
 probe the Kondo destruction effect across the quantum-critical point 
using ARPES and STM measurements is presented.
Particular 
emphasis is placed
on the question of how
to 
distinguish
between the signatures of the initial onset of 
hybridization-gap formation,
which is 
the
``high-energy" physics
to be expected in all heavy-electron systems, and those of
 Kondo destruction,
which characterizes
the low-energy physics and, hence,
 the nature of quantum criticality.
Recent progress 
and possible challenges
in the experimental investigations are surveyed,  the
STM and ARPES spectra for several 
quantum-critical
heavy-electron compounds are compared, and
the prospects for further
advances are outlined.
\end{abstract}

\keywords{heavy-electron quantum criticality,  Kondo destruction, scanning tunneling microscopy, angle-resolved photoemission, hybridization gap}

\maketitle

\tableofcontents
\section{Introduction}

A major objective of quantum materials research
is to  link observable properties to the nature of  quantum mechanical many-body ground state properties and 
to  the characteristics of the excitation spectrum above the ground state. 
In particular, it aims  at understanding and predicting the emergence of novel phases  in terms of a minimal set of variables, 
most notably 
symmetries and broken symmetries
 of the ground state and the ensuing classification of the excitation spectrum. 
The typical energy window commonly involved in the materials of interest  can cover a wide 
range,
 from a few percent of a
 meV to several eV. 
A well-known example  is the high-temperature 
superconductors,
which have stimulated research 
since their discovery more than 30 years ago.

\begin{figure}[t!]
   \centering
   \includegraphics[width=\linewidth]{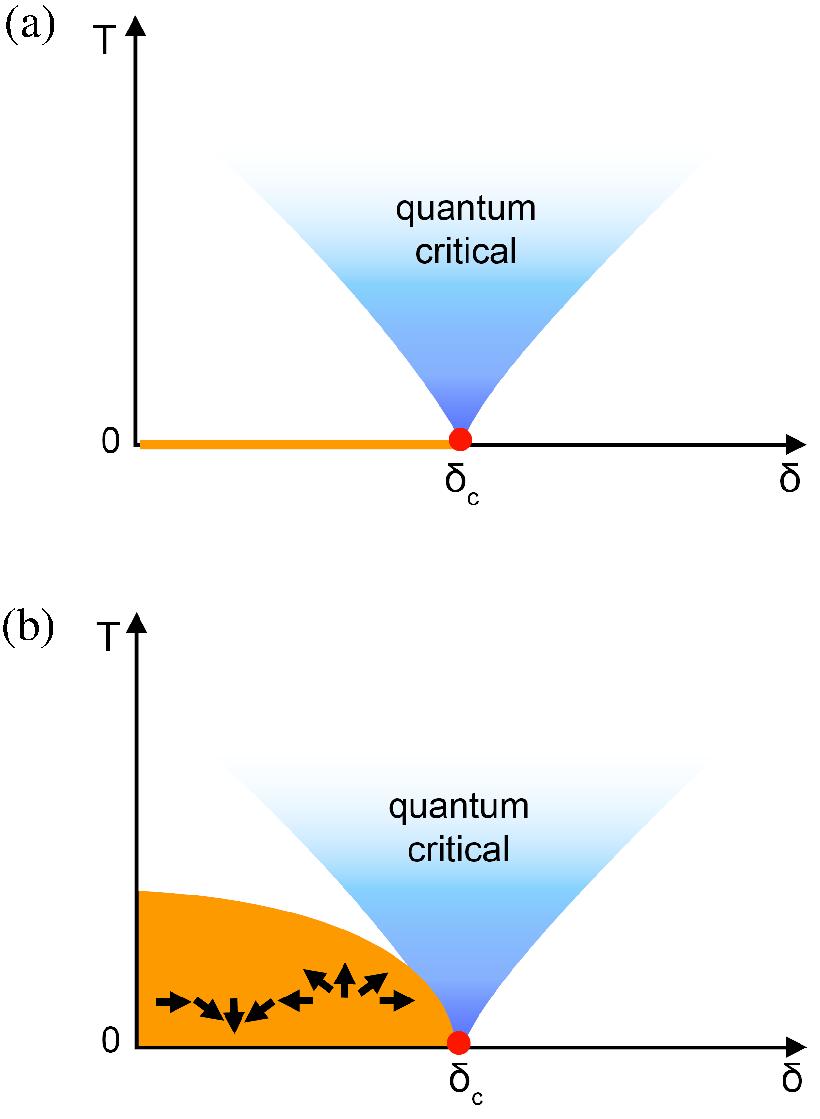}
   \caption{The quantum critical fan. (a) A 
   \qs{continuous}
   quantum phase transition occurs at zero temperature for a critical value ($\delta_c$) of a nonthermal tuning parameter $\delta$. 
   It separates the distinct behaviors of the ground state wavefunction, {\itshape i.e.}, being ordered for $\delta<\delta_c$ (the orange line) and disordered for $\delta>\delta_c$.
   At nonzero temperatures vestiges of the quantum phase transition lead to distinctive scaling behavior in a quantum-critical fan that spreads out of the quantum-critical point and extends up to a problem specific cutoff temperature. Unlike the scaling behavior, 
   the existence of
   a fan of quantum-critical behavior is generic, independent of the nature of the criticality. (b) Frequently, quantum phase transitions occur as order is suppressed and the transition temperature $T_c(\delta)$
   of a classical phase transition vanishes, {\itshape i.e.}, $T_c(\delta\rightarrow \delta_c)\longrightarrow 0$.
   In heavy-electron materials, the most common types of quantum criticality separate antiferromagnetic and paramagnetic phases. Shown here is a type of antiferromagnetic order, indicated by the variation of the magnetic moment density over one wavelength. This Colloquium  explores the potential of 
   \qs{single-particle}
   spectroscopies to distinguish different types of quantum criticality.
   }
   \label{fig:quantcrit}
\end{figure}

The quest for a unified understanding of different classes of quantum materials has led to the notion of quantum-critical points
(QCPs)
as an economic and powerful way of organizing their phase diagrams
\cite{QCNP13,Si.10a,Coleman-Nature,Sachdev-book}.
Such continuous zero-temperature phase transitions not only separate different
ground states
but  
also
give
rise to a characteristic  behavior; this is the quantum-critical fan, 
which can extend to comparatively large energies and temperatures, {\itshape cf.} Fig.\, \ref{fig:quantcrit}. 
Within this fan, universal scaling behavior is expected up to some material-specific high-energy cutoff.
Among the material classes that are currently attracting
particular interest
are the
cuprates, iron pnictides, 
pyrochlore iridates, 
transition metal dichalcogenides, 
and heavy-electron compounds.
An underlying  theme of most if not all of these materials classes  is the tendency of their  charge carriers to localize in response to the large effective Coulomb repulsion experienced by the itinerant degrees of freedom. 
The tendency toward localization gives rise to the bad-metal behavior of these materials.

In  heavy-electron compounds, which most commonly are based on Ce, Yb, and U,
the primary degree of freedom
is
the $f$ electron. In the lanthanide-based materials, the 4$f$ electron is localized close to the ionic core
as a result of atomic physics and thus  has a characteristic energy of the order eV. 
For the same reason,  the wave-function overlap between the 4$f$ orbitals and the band  (or $c$) electrons, 
{\itshape i.e.}, the hybridization,  is typically small. 
As a result, the 4$f$ electron appears localized at high temperatures or energies in the entire range of phase space 
as long as the valency of 
the lanthanide ion remains near its localized limit. 
In this regime
each 4$f$ electron contributes 
\qs{an}
amount $\sim \ln{N_f}$ to the entropy, where $N_f$ is the angular momentum degeneracy. $N_f$ is affected by spin-orbit coupling and the crystal electric fields but as long as $N_f>1$, the spin entropy remains macroscopically large. 

Similar arguments in principle apply to actinide-based heavy-electron compounds \cite{Fisk.85}. In contrast to their  $4f$ counterparts,  $5f$ orbitals  are substantially less localized. As a result, the associated heavy-electron bands are more dispersive, $f-c$ hybridization is stronger and crystal electric fields 
are less well defined. Collectively, these properties frequently lead to more complex behaviors compared to Ce- or Yb-based intermetallics \cite{Lawrence.11}, and so we  use the lanthanide-based heavy-electron materials as exemplary of the essential physics.
As temperature is lowered and the ground state is approached, the spin entropy associated with the localized $4f$ ($5f$) electron needs to be quenched. 
Evidently, the system possesses several options for releasing this 
entropy,
 which 
 lead to different ground states.
 At zero temperature, the system can transition from  one ground state to another upon changing coupling constants in the Hamiltonian. At values of these coupling constants where the ground state 
energy
  is nonanalytic, the system undergoes a quantum phase transition.
Experiments,
however, are performed at 
nonzero temperatures. 
The challenge then is how to distinguish the approach to different ground states with only a limited, intermediate temperature window accessible to experiment. This task is made even more difficult
 given that high-energy properties are
 largely insensitive to the  changes in the coupling constants that take a system through different ground states.

The primary tools for 
exposing
 the underlying physics that accompanies the entropy release as the temperature or energy is lowered 
include
spectroscopic methods that can trace excitations over some energy range of interest.
For example,
spin excitations can be probed with the help of inelastic neutron scattering. 
Among  the various spectroscopic techniques,
angle-resolved photoemission spectroscopy (ARPES) and scanning tunneling microscopy or spectroscopy (STM) stand out as these allow one to most directly trace properties of the 
\qs{single-particle}
Green's function, the basic building block in almost every many-body theory. 

We 
survey and compare
recent ARPES and STM experiments performed on quantum-critical heavy-electron compounds that are located close to ground state instabilites at the border of magnetism. In particular, we focus on how critical 
Kondo destruction
\cite{Si.01,Coleman.01}, 
{\itshape i.e.}, the breakdown of Kondo entanglement at zero temperature right at the onset of magnetism, is reflected in ARPES and STM data at elevated temperatures.

This Colloquium  is organized as follows. After a brief introduction 
of
quantum criticality in heavy-electron systems, we 
recapitulate
 the relation between APRES and STM measurements and their 
 link  with the single-particle Green's function. We then
discuss
 recent STM measurements on YbRh$_2$Si$_2$, a heavy-electron antiferromagnet that features a  
Kondo destruction
QCP
 as a function of the applied magnetic field, before turning to high-resolution ARPES measurements on the cerium-115 family that consists of Ce$M$In$_5$ ($M$=Co,Rh,Ir). We close with an outlook on current challenges and future directions.
 To facilitate reading, each section
 (\ref{sec:QCP}-\ref{sec:115}) ends with a brief summary of the salient points discussed in the section.
\section{Quantum Criticality}
\label{sec:QCP}

Quantum phase transitions occur at zero temperature and like their finite temperature counterparts, they can be either first order or continuous \cite{Sachdev-book,HvL.07,Gegenwart.08,Si.10a}.
In contrast to the finite temperature case where thermal fluctuations drive the transition, quantum fluctuations, encoded already at the Hamiltonian level, are responsible for the occurrence of a quantum phase transition.
A classical transition can be accessed by varying the temperature through a 
\qs{threshold}
 value $T_c$, while the zero-temperature transition is approached by tuning  a nonthermal control parameter, denoted $\delta$ in Fig.\, \ref{fig:quantcrit}, to its critical value ($\delta_c$).
 If the transition is continuous, characteristic, critical scaling ensues in its vicinity which reflects the singular 
 \qs{correlations}
 of the ground state wavefunction at $\delta_c$. At nonzero temperatures, this singular behavior 
 leads to the quantum-critical fan in which characteristic behavior is observed in various quantities below a system-specific cutoff energy: see Fig.\, \ref{fig:quantcrit}.
 
In stoichiometric heavy-electron compounds containing Ce or Yb elements,
4$f$ electrons in a  partially filled 4$f$ shell  are strongly correlated, provided the Ce or Yb ions possess a valence close to +III. 
The spin-orbit interaction and the crystal electric field generated by the ligands surrounding the Ce or Yb ion 
in the crystalline environment reduce the degeneracy of the 4$f$ shell. 
Most commonly, the lowest lying atomic 4$f$ levels correspond to a Kramers doublet.
As a result, the 4$f$ electrons
behave as a 
lattice of effective spin-$1/2$ local moments. 
This leads to an effective description in terms of
the Kondo lattice Hamiltonian:
\begin{eqnarray}
H_{\rm KL}=
H_0
 +
 \sum_{ ij } I_{ij}
{\bf S}_i \cdot {\bf S}_j
+  \sum_{i} J_K {\bf S}_i \cdot {\bf s}^c_i \,,
\label{eq:KM}
\end{eqnarray}
where $H_0= \sum_{ {\bf k},\sigma} \varepsilon _{\bf k}
c^{\dagger}_{ {\bf k} \sigma}
c^{\phantom\dagger}_{{\bf k} \sigma} $ describes the conduction electrons
\sk{with dispersion $\varepsilon_{\bf k}$}.
  The  \sk{Ruderman-Kittel-Kasuya-Yosida (RKKY)}
   interaction $I_{ij}$ \sk{acting between the local moments $\bf S$ at site $i$ and $j$} and the Kondo coupling $J_K$ \sk{(between the local moments and the conduction electron spin density ${\bf s}^c$)} typically are antiferromagnetic, {\itshape i.e.}, $I_{ij}>0$, $J_K>0$.
The competition between these two types of interactions lies at the heart of the microscopic 
physics for heavy-electron systems \cite{doniach1977kondo}.
\begin{figure}[t!]
   \centering
   \includegraphics[width=\linewidth]{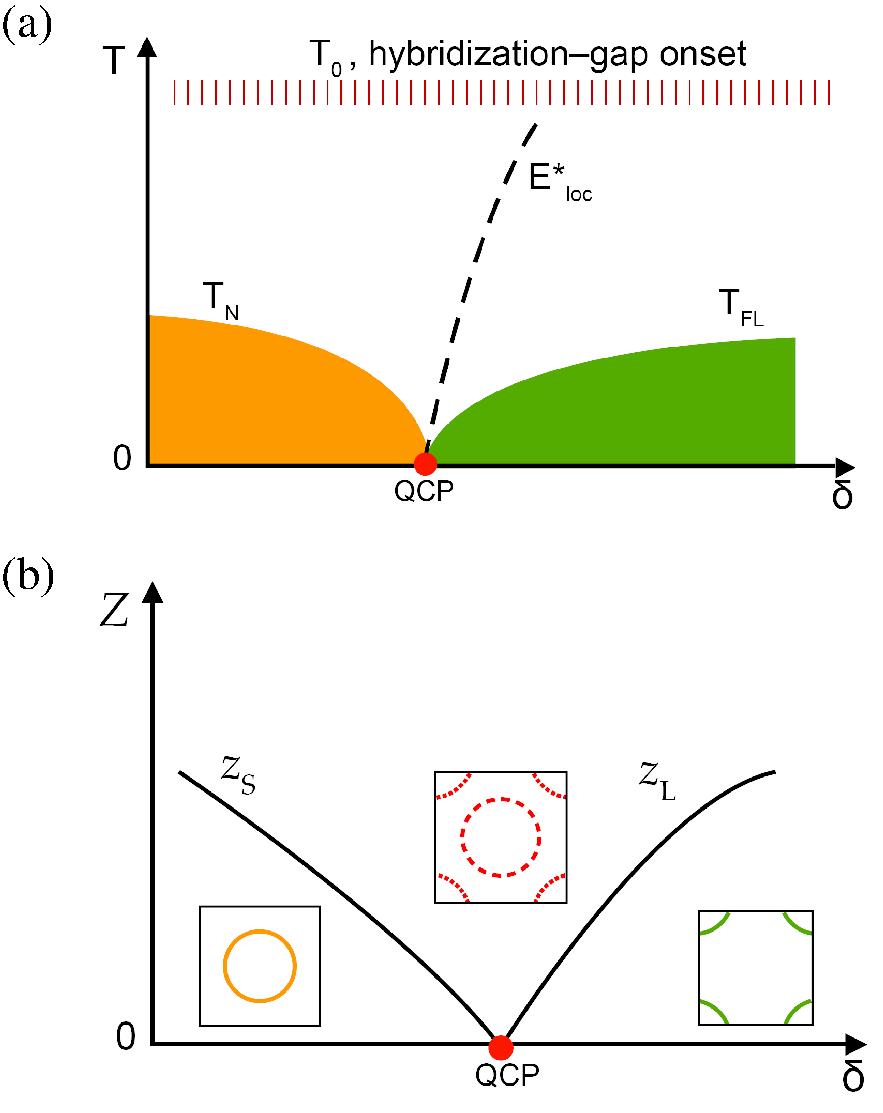}
   \caption{
   Basic concepts of 
   Kondo destruction
    quantum criticality. (a) Local quantum criticality with Kondo destruction under the variation 
    of the control parameter $\delta$.
Here $T_0$ is a high-energy scale that describes the initial onset of dynamical Kondo correlations  and that smoothly 
evolves across the QCP $\delta_c$.
This high-energy scale is reflected in the onset of 
hybridization-gap formation.
The low-energy physics is described in terms of $T_N$ and $T_{\rm FL}$, which are, respectively, 
the temperatures for the N\'eel transition and the crossover into the paramagnetic Fermi liquid state.
This phase diagram  also involves the Kondo destruction energy scale $E_{\mathrm{loc}}^*$,
which
characterizes the Kondo destruction.
The $E_{\mathrm{loc}}^*$ line
divides the phase diagram in terms of the flow of the system toward
either the Kondo screened or the 
Kondo destruction
ground state. In the conventional model of spin-density wave quantum criticality, the line $E^{*}_{\mbox{\tiny loc}}$ extrapolates to zero temperature 
 in the ordered phase so that the Fermi surface already is large
  before reaching the QCP with increasing $\delta$ and evolves smoothly across the QCP \cite{Coleman.01,Si.01}. Adapted from Ref. \cite{Stockert.12}.
(b):  The small (left) and large (right) Fermi surfaces and the associated quasiparticle 
    weights $z_S$ and $z_L$ that are discussed in Sec. \ \ref{sec:ARPES-STM}.
The fluctuating Fermi surfaces
   (middle) are associated with the QCP. Adapted from \cite{Pfau.12}.
   }
   \label{fig:lqcp}
\end{figure}

In the heavy-electron compounds described by the Kondo lattice Hamiltonian, Eq.\, (\ref{eq:KM}), a QCP may arise from tuning the ratio of RKKY to Kondo interactions, which
is parametrized by the nonthermal control paramater $\delta \equiv T_K^0/I$.
Here, the Kondo scale (for the $N_f=2$ case) is  $T_K^0 \approx \rho_0^{-1} \exp \left (-1/\rho_0 J_K \right )$, with $\rho_0$ being 
the density of states of the conduction electrons at the Fermi
energy, whereas
$I$ parametrizes the RKKY interaction.
This RKKY exchange interaction between  the localized moments is mediated by the conduction electron spin density. It is perturbatively generated from the Kondo coupling term $\sim J_K$, resulting in $I_{ij}(J_K)$. In Eq.\,(\ref{eq:KM}), we added $I_{ij}$ as an independent exchange interaction to facilitate the discussion of the phase diagram, because tuning the ratio between the explicit $I_{ij}$ and $J_K$ is more convenient. 
It accesses the quantum phase transition that otherwise would have been induced in the tuning of $\delta$ through the variation of the ratio of $J_K$ to the conduction electron bandwidth $2D \sim 1/\rho_0$.
Formally, one may think of Eq.(1) as arising  from a more complete starting Hamiltonian through the process of integrating out  additional conduction electron degrees of 
freedom; this procedure 
 results in
the explicit
$I_{ij}$ term in the effective Hamiltonian of Eq.(1).
One needs to be sure that there is no double counting of the explicit and generated contributions to $I_{ij}$, and this can be
consistently done in practice.
For a technical discussion of this point see  \cite{Si.05}.

On the paramagnetic side, 
the ground state 
is characterized by the 
   amplitude of the {\itshape static} Kondo singlets that are
   formed between the local moments and conduction electron spins \cite{Hewson}.
   For a 
   Kondo destruction
   QCP, 
   this static
   Kondo singlet amplitude is continuously suppressed
   when
   the system 
   approaches the QCP from 
   the paramagnetic side \cite{Si.01,Zhu.03,Si.14}. 
   
As illustrated in 
Fig.\,\ref{fig:lqcp}(a),
 the Kondo destruction energy scale $E_{loc}^*$ goes to zero as the 
control parameter $\delta$ approaches
the
 QCP at $\delta_c$ from the paramagnetic side,
and the antiferromagnetic  order sets in when $\delta$ goes across 
$\delta_c$.
The Kondo destruction goes beyond the Landau framework of quantum criticality.
The latter is based on order-parameter fluctuations, which  in the present 
context of antiferromagnetic 
heavy-electron systems is  referred to as  a spin-density-wave
(SDW) QCP \cite{Hertz.76,Millis.93,Moriya}.
It arises when $E_{loc}^*$ stays nonzero when decreasing $\delta$ to $\delta_c$
and approaches zero only inside the ordered regime at $\delta<\delta_c$.
In this case, the asymptotic quantum-critical behavior at energies below $E_{loc}^*(\delta_c)$
is the same as in the type of phase diagram shown in Fig.\,\ref{fig:quantcrit}(b), 
where  $E_{loc}^*$ is not part of the critical physics.

The Kondo destruction gives rise to a dynamical spin susceptibility which displays unusual scaling 
at the QCP \cite{Si.01,Si.14}. This includes a fractional 
exponent \cite{Grempel.03,Zhu.03,Glossop.07,Zhu.07}
in the singular dependence on frequency ($\omega$) and temperature ($T$),
and $\omega/T$ scaling.
These 
features 
 have in fact been observed by inelastic neutron scattering measurements on the $5f$ electron system UCu$_{5-x}$Pd$_x$ \cite{Aronson.95} and the $4f$ electron-based metal CeCu$_{6-x}$Au$_x$ \cite{Schroder2}. 
 
For CeCu$_{6-x}$Au$_x$ at its critical doping $x_c\approx 0.1$,
the exponent in the $\omega/T$ scaling analysis \cite{Schroder2}
was found to be $\alpha=0.75(5)$, which compares
well with the value $\alpha=0.72$-$0.78$ calculated at
 the Kondo destruction QCP \cite{Grempel.03,Zhu.03,Glossop.07,Zhu.07}.
In the case of a standard  SDW QCP, no such $\omega/T$ scaling is expected as this QCP is described by a Ginzburg-Landau functional above its upper critical dimension
\cite{Hertz.76,Millis.93,Moriya}.

In the single-particle excitations, the collapse of $E_{loc}^*$ implies a sudden reconstruction of the Fermi surface  across the QCP. 
To contrast this
picture
with the more traditional scenario of an  SDW
 transition \cite{Hertz.76,Millis.93,Moriya}, 
where critical fluctuations are tied to nesting properties of the Fermi surface, we refer to quantum criticality exhibiting critical Kondo destruction as  {\itshape local quantum criticality} \cite{Coleman.01,Pepin.07,Senthil.04,Si.01}. 
 At zero temperature:
\begin{itemize}
\item For $\delta > \delta_c$, the Fermi surface is large and is given by the combination of the 4$f$  and conduction electrons.
A nonzero amplitude of the static Kondo singlet
\qs{specifies}
a
	{\itshape Kondo screened} ground state.
\qs{Here, Kondo resonances appear in the excitation spectrum, reflecting 
the entanglement between the  4$f$ moments  and the conduction electrons
in the ground state.}
 The Kondo effect is responsible 
for the large mass enhancement and a small quasiparticle weight $z_L$: see 
Fig.\,\ref{fig:lqcp}(b). There is a small gap for the single-particle excitations
\qs{at the wavevectors where the small Fermi surface would have resided.}

\item  For $\delta < \delta_c$, the Fermi surface is small as determined by the conduction electrons alone.
This is because, when the amplitude of the static Kondo singlet vanishes, there is no longer 
	a well-defined Kondo resonance.
	We refer to this state as a {\itshape Kondo destruction} ground state.

\item At the QCP,  single-particle excitations are gapless and have a non-Fermi-liquid form, at
both  small and large Fermi surfaces. 
\end{itemize}

\subsection{High-energy excitations, temperature evolution, and mass enhancement}

\begin{figure}[t!]
\centering
\includegraphics[width=\linewidth]{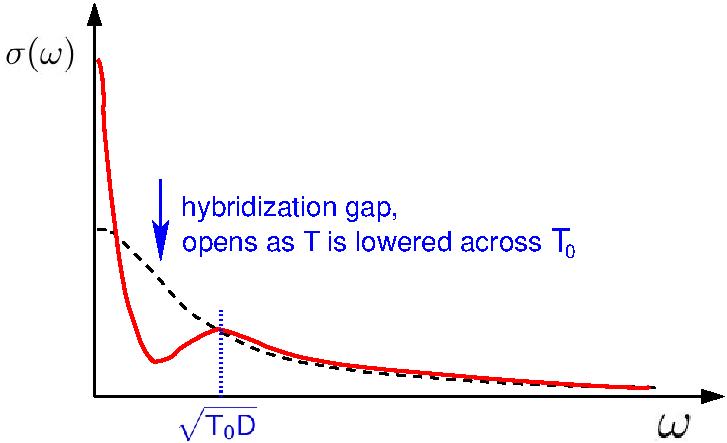}
\caption{
Sketch of the optical conductivity $\sigma(\omega)$ for temperatures well above (dashed black line) and well below (continuous red line) the crossover temperature
 scale $T_0$.
 Here the lowering of temperature through $T_0$ is accompanied by 
 the 
 onset of the hybridization gap.
 The characteristic frequency scale for the hybridization gap is 
 $\sqrt{T_0D}$ (vertical dotted blue line),
 where $D$ is an energy scale of the order of  the conduction electron bandwidth.
 At low energies, {\itshape i.e.}, for $\omega\ll T_0$ and $T\ll T_0$, 
 and sufficiently far away from quantum criticality ({\itshape i.e.} $\delta <\delta_c$ or $\delta>\delta_c$), 
 a pronounced Drude peak reflects the mass enhancement in the Fermi liquid regimes that surround the QCP
in the phase diagram: see Fig.\ \ref{fig:lqcp}(a).
The behavior of $\sigma(\omega)$  at high energies, including the hybridization gap, is a generic feature of heavy-electron systems and  is seen throughout the high-energy part of the phase diagram of 
Fig.\ \ref{fig:lqcp}(a).
 }
 \label{fig:optcond}
\end{figure}

Figure \ref{fig:lqcp}(a) 
also contains a high-energy scale $T_0$ which 
describes 
{\itshape the initial onset}
 of dynamical Kondo correlations.  This scale is  generally affected by the presence of higher crystal electric field doublets (or quartets) that together form the 4$f$ multiplet \cite{Cornut.72,Kroha.03,Chen.17,Pal.19}.
 It is important to note that this scale smoothly evolves across the QCP
 at
  $\delta_c$.
The
 development of the hybridization gap is associated with the initial onset of dynamical Kondo correlations, 
as illustrated in Fig.\,\ref{fig:optcond},
and will appear
on both sides of $\delta_c$.

For $\delta > \delta_c$, the temperature evolution of the physical properties reflects the flow of the system toward
the
Kondo screened
ground state. For instance,
the initial onset of dynamical Kondo correlations  results in the Kondo screened ground state;
the single-particle excitations 
develop into fully coherent 
heavy quasiparticles at the large Fermi surface as the temperature is lowered below
$T_{\rm FL}$,
the crossover temperature into the paramagnetic Fermi liquid state.

For $\delta < \delta_c$, the initial 
onset
of dynamical Kondo correlations
still takes place, even though it does {\itshape not}, in the end,  
	lead to a well-defined Kondo resonance and 
	the Kondo-singlet amplitude vanishes
	in the ground state. Still,
as the temperature is further lowered,
vestiges of the Kondo effect will be observed at any nonzero temperature.
In particular, the effective mass is a dynamical quantity, measuring the dispersion of the Landau quasiparticles,
and is
enhanced through the dynamical Kondo effect;
further discussions of this point have been given by \cite{Cai.1904,Si.14}.

\subsection{Isothermal evolution at low temperatures}

The distinction between the two sides of $\delta_c$ can be sharply made at low temperatures, where well-defined quasiparticles 
reside at the small Fermi surface for $\delta<\delta_c$ and at the large Fermi surface for $\delta>\delta_c$. At zero temperature, 
\qs{a sudden reconstruction of the Fermi surface appears}
as $\delta$ passes through $\delta_c$. At nonzero but low temperatures, this becomes a crossover. 
The crossover width increases with increasing temperature. When the crossover width becomes
 large, the difference between the two sides  becomes ambiguous. 
 We  illustrate this point next, especially through the experiments carried out on YbRh$_2$Si$_2$.
 
 \subsection{Further considerations}
In the Kondo destruction description, the {\it static} Kondo effect is suppressed in 
the antiferromagnetic phase at $\delta<\delta_c$.
However, dynamical Kondo singlet correlations remain at nonzero frequencies in this regime. 
They lead to the development of 
	$4f$-electron spectral weight near the Fermi energy, which we  refer to as Kondo resonance-like features.
The dynamical Kondo effect [see Ref.\,\cite{Cai.1904}] as well as earlier discussions by
\cite{Si.14,Zhu.03}
 still produces a large mass enhancement and a small quasiparticle weight $z_S$: see 
Fig.\,\ref{fig:lqcp}(b).
There is a small gap for the 
single-particle excitations at 
\qs{the wavevectors where the large Fermi surface would have developed.}

The inelastic neutron scattering result on 
CeCu$_{6-x}$Au$_{x}$ ($x_c \approx 0.1$) \cite{Schroder2}
has been confirmed by the recent
inelastic neutron scattering measurements \cite{Poudel.19}
in CeCu$_{6-x}$Ag$_{x}$ ($x_c \approx 0.2$).
When analyzing the data in terms of the one-component form, as arising in the 
Kondo destruction description, 
Ref.~\onlinecite{Poudel.19} found a similar form of 
 $\omega/T$ scaling with a similar value for the 
 critical exponent $\alpha=0.73(1)$. 
 Ref.~\onlinecite{Poudel.19} also analyzed 
the data 
 in terms 
of a multicomponent spin fluctuation spectrum with one of the weaker
components conforming to the expectation of an SDW QCP. 
However, thermodynamic singularities have provided evidence for 
 the one-component description \cite{Grube.17}.

\subsection{Summary of Sec.\,\ref{sec:QCP}}
For heavy-electron metals,
the Landau form of quantum criticality 
corresponds to 
an SDW QCP. A new type of quantum criticality has been advanced
in the form of a Kondo destruction (local) QCP. It goes beyond the Landau framework
in that the critical destruction of Kondo entanglement characterizes the 
physics beyond the slow fluctuations of the magnetic order parameter.
The Kondo destruction is characterized by a new energy scale
$E_{loc}^*$ vanishing at the QCP as illustrated in Fig.\,\ref{fig:lqcp}(a);
a sudden reconstruction from large to small Fermi surface  as the system
is tuned from the paramagnetic side through the QCP, along with a vanishing quasiparticle weight on approach of the QCP from both sides, as illustrated in 
Fig.\,\ref{fig:lqcp}(b). 


\section{ARPES, STM, and the Single-particle Green's Function}
\label{sec:ARPES-STM}

The unusual $\omega/T$ scaling of the dynamical spin susceptibility  sets apart the QCP featuring  
critical Kondo destruction 
 from the more traditional QCP 
 based on the Landau framework of order-parameter fluctuations.
It means that the dynamical susceptibility $\chi(\omega,T)$, in the regime where 
\qs{the critical singularities dominate,}
 can be scaled to depend on $\omega$ or $T$ only through
 the combination $\omega/T$.
Such scaling
has been  observed in  CeCu$_{6-x}$Au$_x$ at its 
antiferromagnetic
QCP \cite{Schroder1} and indicated for  YbRh$_2$Si$_2$ \cite{Friedemann.10}. 
Recent measurements of the optical conductivity in thin films of YbRh$_2$Si$_2$  have demonstrated 
a singular response in the charge sector with an $\omega/T$ scaling  \cite{Prochaska.18}. 
A scaling form of this kind for both the optical conductivity and dynamical spin susceptibility  is strongly suggestive of the presence of  $\omega/T$ scaling in the single-particle excitations encoded in the 
\qs{single-particle}
Green's function.

\subsection{The 
\qs{single-particle}
Green's function}
\label{subsec:2-1}

This Green function can quite generally be cast into the form
\begin{equation}
G(\omega,{\mathbf{k}},T)=\frac{1}{\omega-\varepsilon_{\mathbf{k}}-\Sigma(\omega,{\mathbf{k}},T)},
\end{equation}
where $\varepsilon_{\mathbf{k}}$ is the bare electron dispersion and the proper self-energy $\Sigma(\omega,{\mathbf{k}},T)$ encodes the effects of electron-electron interaction. In a Fermi liquid, this function can be decomposed into two parts,
\begin{equation}
G(\omega,{\mathbf{k}},T)=G_{\mbox{\tiny coh}}(\omega,{\mathbf{k}},T)+G_{\mbox{\tiny incoh}}(\omega,{\mathbf{k}},T),
\end{equation}
where the incoherent part is nonsingular close to the Fermi surface while the coherent part $G_{\mbox{\tiny coh}}$ near $E_F$ describes the quasiparticle contribution and assumes the form
\begin{equation}
G_{\mbox{\tiny coh}}(\omega,{\mathbf{k}},T)=\frac{z_{}}{\omega-v_{F}(k-k_F)+i\Gamma(\omega,T)},
\end{equation}
where $z$ is the quasiparticle weight and $v_F$ is the Fermi velocity. The lifetime of a quasiparticle is given by the inverse of the decay rate $\Gamma$.
The amplitude of the static Kondo screening previously discussed is related to a pole in the self-energy, which  in the Fermi liquid regime can at sufficiently low $\omega$ and $T$ be written as
\begin{equation}
\Sigma(\omega,\mathbf{k},T)=\dfrac{a}{\omega-b}+\delta \Sigma(\omega,\mathbf{k},T),
\end{equation}
where $a$ and $b$ are parameters 
that capture the strength of Kondo screening and the energy of the Kondo resonance,
respectively.
The pole in $\Sigma$ shifts the Fermi momentum from its initial, ''small'' value to a new, ''large'' value as long as $a\neq 0$.
\qs{In the SDW QCP case, $a\neq 0$ on either side of the critical point. 
For the local QCP, by contrast, Kondo screening is critically destroyed; correspondingly, 
on the antiferromagnetic side, 
 $a=0$ and the Fermi surface is small.
 } 
 
The Hall effect turns out to be a particularly useful quantity in this context as it is a measure of the carrier density on either side of the QCP. 
This is a consequence of the Fermi liquid nature of the two phases separated by the QCP,
in which the Hall coefficient is  completely determined by the
renormalized dispersion of the
   \qs{single-particle}
excitations, to the leading order of elastic scattering (quenched disorder) when it is 
\qs{nearly isotropic. In other words, here, the Hall coefficient}
is independent of  the quasiparticle weight $z$ or any Landau parameters,
regardless of the strength of electron-electron (and electron-phonon) interactions.
This can be seen through the kinetic equations of a Fermi liquid or
using the Kubo formalism  \cite{Betbeder.66,Kohno.88} 
 and 
related Feynman diagrammatic means \cite{Khodas.03}.

The dynamical spin susceptibility $\chi(\omega,\mathbf{k},T)$ and also the optical conductivity $\sigma(\omega,T)$ can be written as convolutions of the Green's function with itself and specific vertex functions. On the other hand, ARPES and STM measurements depend directly on $G(\omega,{\mathbf{k}},T)$.  
   \qs{Single-particle}
    spectroscopies are thus, at least in principle,  particularly useful in  distinguishing between the two types of quantum criticality.

	\subsection{ARPES and STM}
	\label{subsec:2-2}

\begin{figure}[htb]
\centering
\includegraphics[width=0.95 \linewidth]{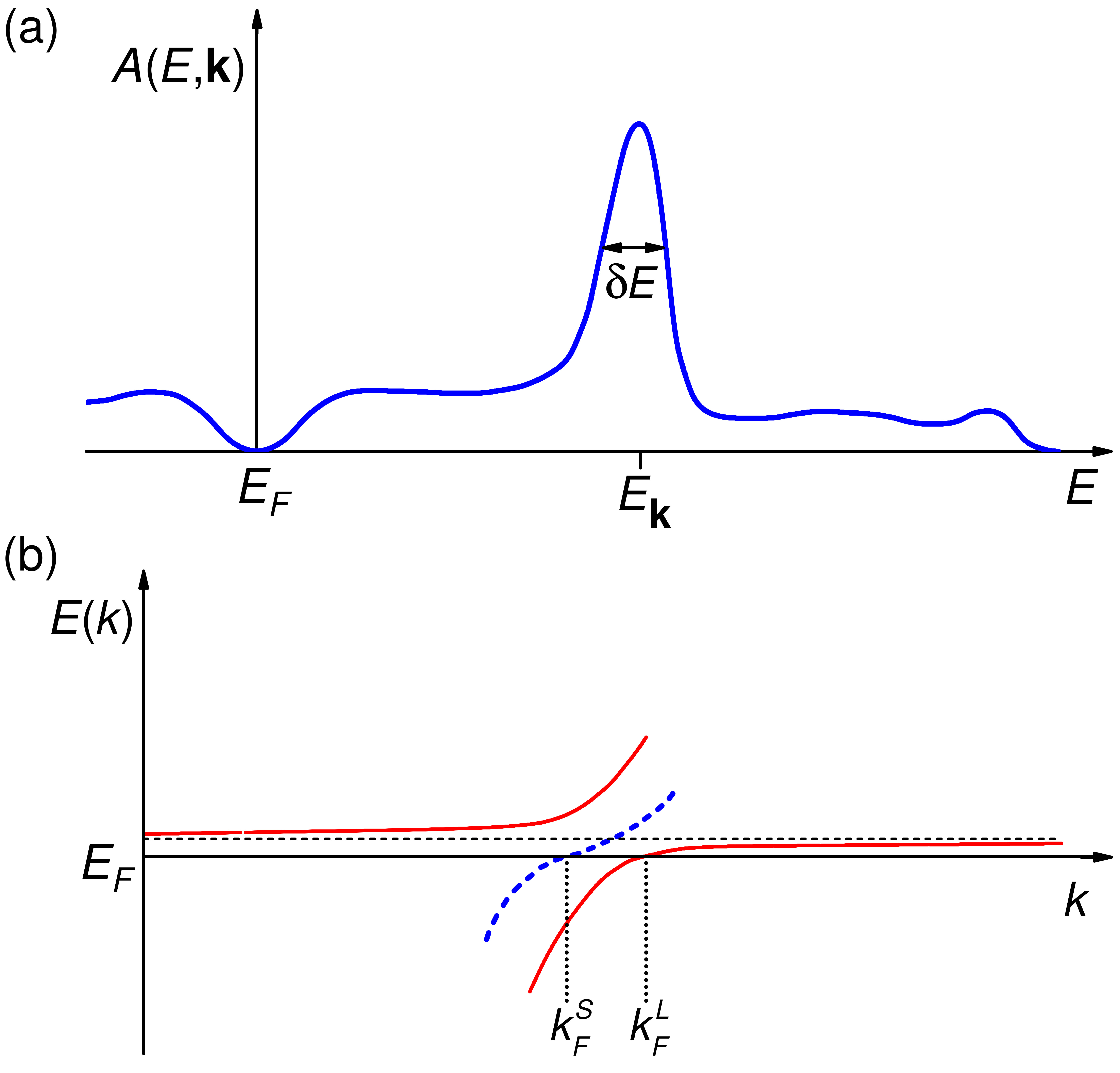}
\caption{Electronic characteristics of the Fermi liquid state of a Kondo lattice. (a) Spectral density of a Fermi liquid: The quasiparticle pole at $\epsilon_k$ has a characteristic width $\delta \epsilon$ that increases with the distance from the Fermi energy $E_F$ as 
\qs{$\delta E \sim |E_\mathbf{k}-E_F|^2$. }
A similar broadening occurs due to finite-temperature effects. The incoherent part of
\qs{$A(E,\mathbf{k})$}
vanishes at $E_F$.
 (b) Quasiparticle dispersion in the Fermi liquid to either side of the QCP:
$k_F^L$ and $k_F^S$ refer to large and small Fermi surfaces, respectively. 
Across a Kondo destruction QCP, the one-electron spectrum is gapless at $k_F^L$ and develops a small gap at $k_F^S$ 
for $\delta>\delta_c$,  and the converse is valid for $\delta<\delta_c$.
The flattening of the dispersion near $k_F^S$ for $\delta<\delta_c$ (dashed blue curve)
reflects the
effective mass enhancement 
due to the 
dynamical
Kondo effect.}  
 \label{fig:sketch}
\end{figure}

ARPES and STM measurements probe the single-particle spectrum and thus give access to the  spectral function 
\qs{$A(E,{\mathbf{k}})=-(1/\pi) \mbox{Im} G(\omega=E+i0^+,{\mathbf{k}},T)$.}
Although the single-particle Green's function appears in the theoretical description of 
ARPES and STM, both spectroscopic techniques are complementary.
While ARPES directly  
probes the single-particle excitations as a function of energy and momentum, STM measures a conductance that is local in real space. Both methods are surface sensitive, albeit to different degrees. Furthermore, through variation of the photon energy, the bulk sensitivity of ARPES can be enhanced. By construction, ARPES probes only the occupied part of the single-particle excitation spectrum, which, especially at low temperatures, leads to a sharp cutoff at the Fermi energy \cite{Huefner}. ARPES therefore measures only part of the full spectral function, {\itshape i.e.},  the imaginary part of the retarded Green's function below the Fermi energy. A sketch of the spectral function of a Fermi liquid is shown in 
Fig.\,\ref{fig:sketch} (a). 
It consists of  contributions from the quasiparticle pole and an incoherent background. The quasiparticle pole contributes a factor $z$ to the total area beneath the spectral function, while the incoherent background contributes $1-z$ times the total area. $z$ is commonly called the wave-function renormalization factor and  it is inversely proportional to the quasiparticle mass in a Fermi liquid. The evolution of $z$ with tuning parameter is plotted schematically in Fig.\,\ref{fig:lqcp}(b) where we see that $z$ vanishes at a local QCP.
The position of the quasiparticle pole as a function of momentum defines the disperion. 
{\itshape Per se}, ARPES is not able to distinguish between the quasiparticle peak and the incoherent part of the spectral function.
Provided the energy and momentum resolution is not a limiting factor, however, the characteristic broadening $\delta \epsilon$ of the quasiparticle peak in energy  
\qs{($\sim |E_{\mathbf{k}}-E_F|^2$,
where $E_{\mathbf{k}}$ is the quasiparticle energy)}
and with temperature ($\sim T^2$) should be discernible in the momentum-distribution curves provided by ARPES.
The weight of the quasiparticle peak,  in principle, could also be extracted based on the total incoherent part. However, since ARPES probes only occupied states, the complete spectral function is inaccessible.  
Although  inverse photoemission is in principle able to probe  states above the Fermi energy, it is 
\qs{limited}
 by  rather poor energy resolution. 

In general, when interpreting ARPES spectra, one needs to keep in mind that in order to
relate  the photoemission intensity to the spectral function,
the one-electron dipole matrix element enters, which generally is unknown. 
In addition, $k_z$ broadening  can be important, where $k_z$ is the component 
of the electron momentum  perpendicular to the surface and depends on the photon energy \cite{Strocov.03,Wadati.06}.

STM, on the other hand, measures a local-in-real-space conductance.
In the linear-response regime, 
the current-voltage characteristics are related to the local density of states (DOS) of the 
material under investigation \cite{Bardeen.61,Tersoff.85}. Therefore,  at low bias voltage and temperature, 
the spatially resolved spectral density  can be obtained. As the applied bias voltage shifts the chemical potential 
at which the local density of states is probed, STM is, unlike ARPES,  not confined to only occupied states. 
It is, however, important to realize that the assumption that the spectral function is independent of the bias
 voltage has to break down at some sample-dependent value of the bias voltage beyond which the tunneling 
 current can no longer be related to the local density of states. Moreover, the properties of the STM tip, {\itshape e.g.}, its DOS, may affect the results.

\subsection{Probing quantum criticality in the Kondo lattice}
\label{subsec:2-3}

One of the strongest diagnostic tools to distinguish the local QCP from the SDW QCP is Hall conductivity measurements, as the local QCP manifests itself by a jump of the Hall coefficient across the QCP. This is a consequence of  the Hall coefficient being inversely proportional to the carrier density (in the isotropic case or, in general, the curvatures of the quasiparticle dispersion on the Fermi surface) while being independent of the quasiparticle weight $z$ in a Fermi liquid. The continuous nature of the local QCP is ensured by the vanishing of the quasiparticle weight from either side of the transition. 
For an SDW QCP, on the other hand, $z$ will remain 
nonzero (except at isolated points on the Fermi surface)
as $\delta$ is tuned through $\delta_c$.

In a Kondo lattice at sufficiently high temperatures, where in first approximation
 the effect of the RKKY interaction can be ignored, 
the single-impurity Anderson model is expected to capture the overall physical behavior. This model is given by
\begin{eqnarray}
\label{eq:Hamiltonian}
H_{\mbox{\tiny AND}}&=&\sum_{\sigma}\varepsilon_{} f^\dagger_{\sigma}f^{}_{\sigma}+
\sum_{\mathbf{k},\sigma}\varepsilon_{\mathbf{k}}c^\dagger_{\mathbf{k},\sigma}c^{}_{\mathbf{k},\sigma} \\
&+&\frac{U}{2}\sum_{\sigma \neq \sigma'}f_{\sigma}^{\dagger}f_{\sigma'}^{\dagger}f_{\sigma'}^{}f_{\sigma}^{} 
+\sum_{\mathbf{k},\sigma}\Big(V_{\mathbf{k}}^{ } f_{\sigma}^{\dagger}c_{\mathbf{k},\sigma}^{}+\mbox{h.c.}\Big) \nonumber
\end{eqnarray}
where  $f^\dagger_{\sigma}$ 
($f_{\sigma}^{}$)
 is the set of local 4$f$ electron creation (destruction) operators of spin projection $\sigma$. The conduction electron operators are $c^\dagger_{\sigma}$ 
 and
 $c_{\sigma}^{}$. The band structure of the conduction electrons is encoded in $\varepsilon_{\mathbf{k}}$, and
the matrix element $V_{\mathbf{k}}^{ }$ that mixes 4$f$ and $c$ electrons is 
 referred to as  the hybridization.
[For the case of the periodic Anderson model
in
 the local-moment limit, with the 4$f$ electron occupancy being close to unity, it reduces to the Kondo-lattice model given in 
Eq.\,(\ref{eq:KM}) when the charge degrees of freedom of the 4$f$ electrons  are projected out \cite{Schrieffer.66,Zamani.16}.]

STM spectra of single-site Kondo problems  
possess the structure of Fano resonances \cite{Fano.61} and depend on the ratio of  tunneling into the Kondo impurity vs tunneling into the embedding host. This ratio is encoded in the so-called Fano parameter. 
Rigorous derivations of the tunneling current and the form of the Fano parameter 
are
 given in \cite{Schiller.00,Ujsaghy.00,Plihal.01}. 
If  tunneling
occurs
 predominantly into
 the
  conduction band 
    the measured  local DOS features the suppression of conduction electron states near the Fermi energy as the Kondo effect develops. 
The first scanning tunneling studies of dense Kondo systems 
appeared about a decade ago
   \cite{Aynajian.10,Schmidt.10,Ernst.10}.
   The pronounced variation  of  STM spectra with the type of surface for Kondo lattice compounds is largely due to variations in the Fano parameter (see {\itshape e.g.}, 
Fig.\,\ref{fig:STM-115}
 for tunneling into
differently terminated surfaces).
This has been explicitly demonstrated based on  mean field and dynamical mean field theory approximations for the Kondo lattice \cite{Dzero.09,Woelfle.10,Figgins.10,Benlagra.11}.
At sufficiently high  temperatures,  however, STM spectra in the vicinity of each Ce moment are expected
 to be similar to those for the  single-ion Kondo case. Kondo screening is a predominantly local phenomenon and 
 thus its onset and evolution are easily probed in real space, {\itshape i.e.}, via STM.
For a study of the single-particle Green's function in the paramagnetic Fermi liquid regime of the Kondo lattice
 far away from any QCP, in the context of photoemission, see Refs.\,\cite{Reinert.01,Costi.02}.
ARPES measurements at similar temperatures, around and above the energy scale $T_0$, provide the band structure $\varepsilon_{\mathbf{k}}$ 
of
the occupied conduction electron states.  A flat band near 
the 4$f$ electron atomic level
$\varepsilon_{}$ [see Eq.\ (\ref{eq:Hamiltonian})], 
which is far from the Fermi energy,
and the formation of a flat band near the Fermi energy 
induced
 by 
 the
 Kondo
 effect
  at each Ce moment reflect the 4$f$ electron spectral weight. This can be enhanced using resonant ARPES,  \cite{Chen.17}.

At sufficiently low temperatures, in the Fermi liquid regime to either side of the QCP at $\delta_c$
[Fig.\ \ref{fig:lqcp}(a)],
\qs{the band structure near the Fermi energy is shown in}
Fig.\,\ref{fig:sketch}(b).
For $\delta<\delta_c$, the small Fermi surface prevails and the band structure is that of the blue dashed  line crossing the Fermi energy $E_F$ at $k_F^S$.
Still, incoherent spectral weight, 
a vestige
of 
 incomplete Kondo screening, 
develops
 but is ultimately gapped near $k_F^L$.
For $\delta>\delta_c$,  the Fermi surface incorporates the $4f$ moments and the Fermi wavevector  changes from $k_F^S$ at high temperatures (without the $4f$ moments) to $k_F^L$ at low temperatures. On this side of the QCP, any spectral weight near $k_F^S$ is due to incoherent single-particle excitations and 
is ultimately gapped. In other words, for
$\delta>\delta_c$ in the Fermi liquid regime the spectral weight near the dashed blue line of  
Fig.\,\ref{fig:sketch}(b) has developed 
a 
small gap at $k_F^S$.
This should in principle be  directly detectable via  ARPES, provided the energy and momentum resolution are sufficiently high, and low enough temperatures can be reached.

On the other hand, the change $k_F^S$ to  $k_F^L$ has only indirect vestiges in real space 
as the Fermi liquid is a momentum-space concept.
The ensuing difficulties when tracing 
   \qs{single-particle}
   excitations in real space  can already  be read off from  
Fig.\,\ref{fig:sketch}(b): The Fermi liquid is described by a low-energy effective theory and is valid only in the vicinity of $k_F$. 
[The spectral function is a more general concept but it only  assumes a form 
\qs{as illustrated}
in 
Fig.\,\ref{fig:sketch}(a) in the Fermi liquid regime.]
Fourier transforming the momentum-resolved spectral function to real space necessarily 
 will sum up  spectral weight outside of the Fermi liquid regime, 
 where
 \qs{the characteristic form of broadening}
 that 
 identifies
 the quasiparticle peak is no longer valid.

One possible way forward  is to perform quasiparticle interference (QPI) experiments to map out the band structure near the Fermi energy \cite{Derry.15,Yazdani.16}. 
We  return to this possibility in Sec.\ \ref{sec:PCO}. 
Another possible way is to perform isothermal STM measurements at low temperatures through the phase diagram connecting $\delta<\delta_c$ with $\delta>\delta_c$. While this by itself does not provide any direct information on the size of the Fermi surface, it was recently demonstrated that such a measurement is 
able to pick up the critical slowing down
at the Kondo destruction energy scale \cite{Seiro.18}, as discussed in the next section \sk{(see Sec.\, \ref{sec:YRS})}.

\subsection{Summary of Sec.\,\ref{sec:ARPES-STM}}
The nature of quantum criticality in heavy-electron metals is manifested in the evolution of the 
single-particle excitations across the QCP.
It is natural to probe this behavior using the ARPES and STM spectroscopies, given that 
they are an established means of studying single-particle excitations in metals.
However, this task is challenging, mostly because 
heavy-electron systems have the distinction that the required energy 
scale is very low. 

For ARPES, this requirement poses a challenge to access the quantum-critical behavior as is the limitation that even the state-of-the-art setups 
cannot yet reach temperatures below about $1$ K.
Still, ARPES should be informative in elucidating
(i) the onset of a hybridization gap,
which represents the high-energy physics for the quantum criticality of heavy-electron
metals [see Fig.\,\ref{fig:lqcp}(a)]; and (ii) the evolution of the 
dynamical Kondo effect as temperature is lowered toward either the antiferromagnetic or paramagnetic
ground state or the quantum-critical regime.

STM spectroscopy has superb energy resolution and can reach low temperatures,
but more demanding setups (such as those suited for QPI) are needed to access the information in the momentum space.
Still, STM probes the single-particle physics in a way that is complementary to ARPES.
In addition, it provides a promising means to probe the isothermal evolution of single-particle
excitations at low temperatures across the Kondo destruction energy scale.

\section{Quantum Criticality in Y\MakeLowercase{b}R\MakeLowercase{h}$_2$S\MakeLowercase{i}$_2$}
\label{sec:YRS}

\begin{figure}[t!]
	\centering
	\includegraphics[width=\linewidth]{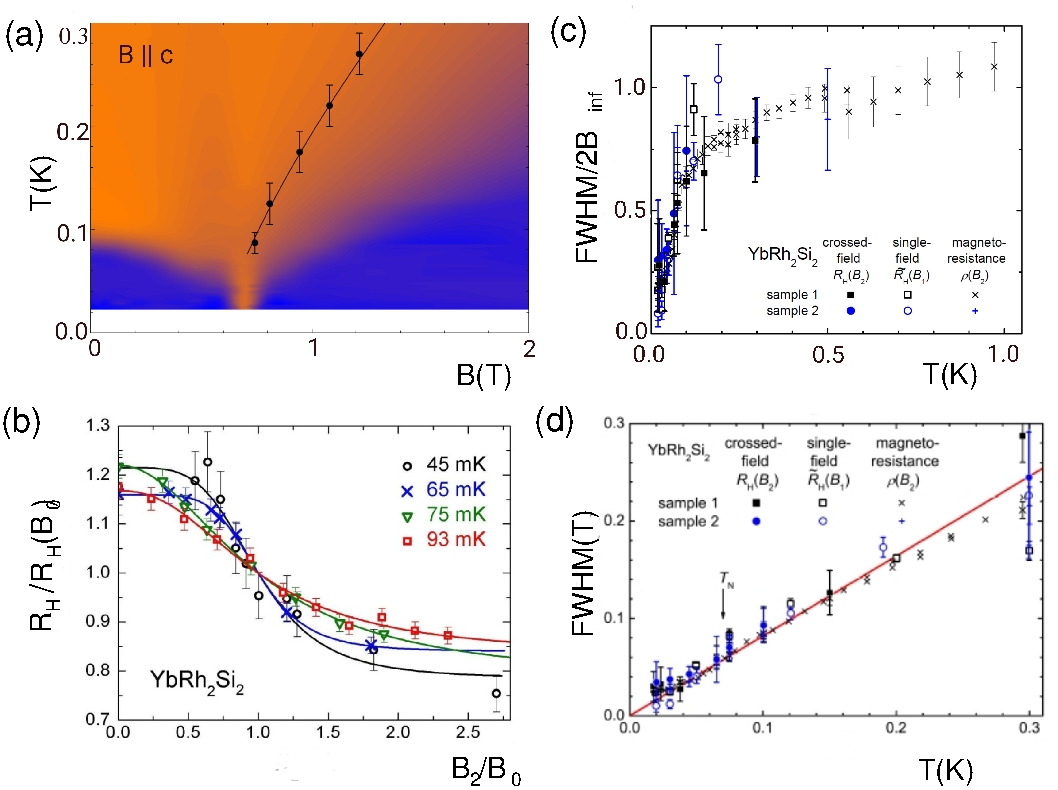}
	\caption{Quantum criticality in YbRh$_2$Si$_2$. (a) The temperature vs field phase diagram of YbRh$_2$Si$_2$  The blue regions mark Fermi liquid behavior, {\itshape i.e.}, $\rho(T)-\rho(0)\sim T^2$, while orange indicates the quantum-critical area of the phase diagram where $\rho(T)-\rho(0)\sim T^x$, with $x\approx 1$. The continuous line in the quantum-critical region is the $E^*$ line as derived from thermodynamic and 
\qs{transport properties \cite{Paschen.04,Gegenwart.07,Friedemann.10}. Adapted from}
\cite{Custers.03}.
		(b) Normalized Hall coefficient across the critical field for different temperatures. The inverse of $R_H$ is a measure of the carrier density. The lower $T$, the sharper is the crossover. At $T=0$ and $B=B_c$, a jump of $R_H$ corresponds to the sudden localization of 4$f$ electrons as $B$ is taken trough $B_c$ from above.
		From \cite{Paschen.04}.
		(c)  Comparison between the isothermal magnetotransport crossover width and the crossover field as specified by the ratio of the FWHM/2 to the crossover inflection field $B_{\rm inf}$. FWHM denotes the full width at half maximum. From \cite{Paschen.16}.
		(d) The ``sharpness" of the crossover: The FWHM vanishes in a linear-in-$T$ fashion indicating a jump of $R_H$ at $B_c$ in the zero-temperature limit. \sk{From \cite{Friedemann.10}.}
	}
	\label{fig:yrs_figure1}
\end{figure}

YbRh$_2$Si$_2$ is a prototype system for local quantum criticality as illustrated by 
its temperature ($T$)-magnetic field ($B$) phase diagram:  
see Fig.\,\ref{fig:yrs_figure1}(a). 
Here the Fermi surface jump and the Kondo destruction energy scale 
have been extensively studied through magnetotransport and thermodynamic measurements.
At a given temperature, the isothermal Hall coefficient 
[Fig.\,\ref{fig:yrs_figure1}(b)]
and other transport and thermodynamic
quantities display a rapid crossover \cite{Paschen.04,Gegenwart.07,Friedemann.10}.

From these measurements, a
$E^*(B)$ line is thus specified in the phase diagram. This line relates to each $T$ a $B^*$ scale: $B^*(T)\geq B_c$ with  $B^*(T=0)= B_c$.
The full-width at half-maximum (FWHM) of the crossover [Fig.\,\ref{fig:yrs_figure1}(b)] extrapolates to zero in the zero-temperature limit 
[Fig.\,\ref{fig:yrs_figure1}(d)],
which implicates a jump of the Fermi surface across the QCP.
It follows that, in the low-temperature limit, at $B<B_c$, the Fermi surface is small.

\begin{figure}[t!]
\centering
\includegraphics[width=0.85 \linewidth]{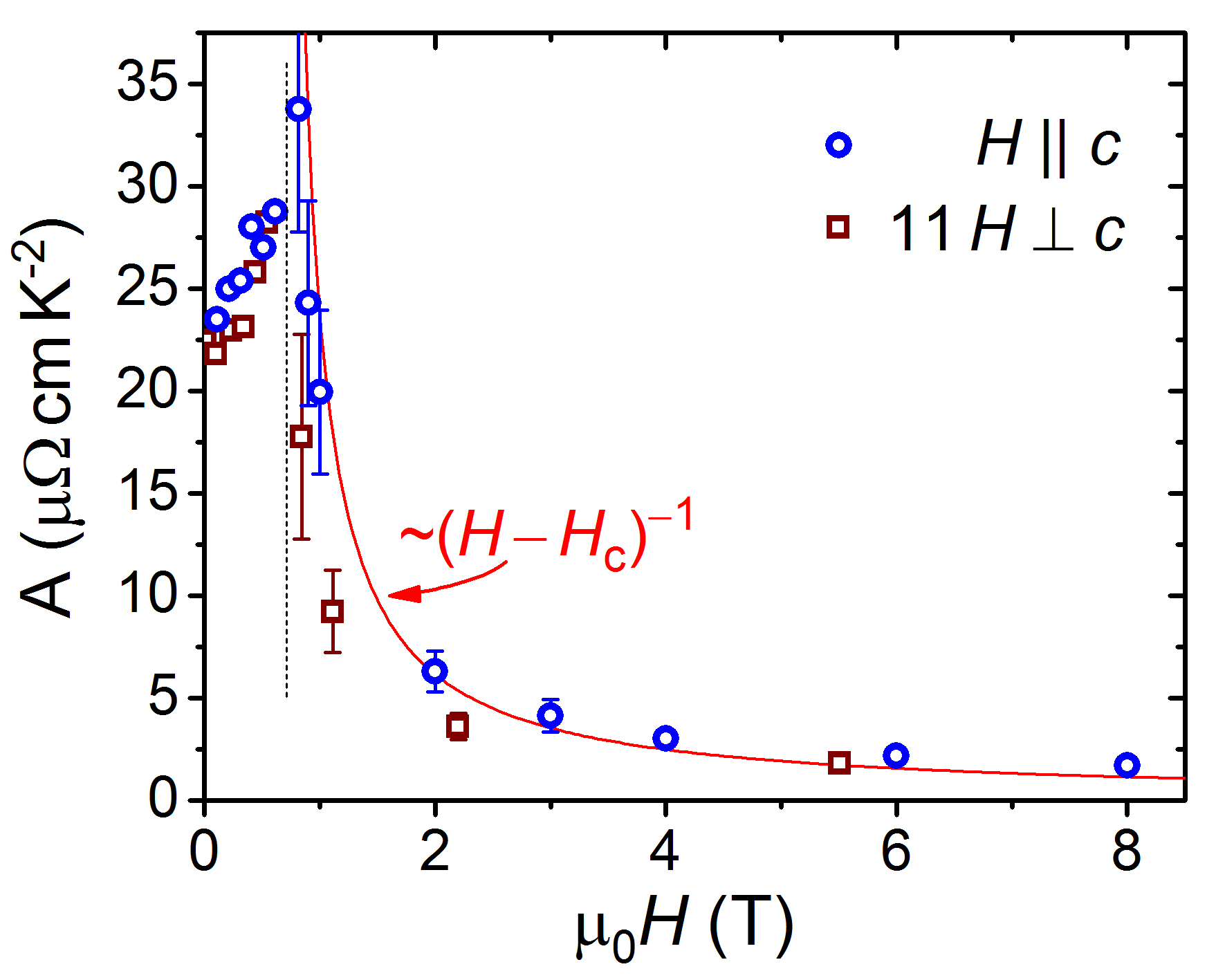}
\caption{Divergence of 
\qs{the $T^2$ coefficient $A$ of the resistivity}
at the QCP in YbRh$_2$Si$_2$. 
\qs{Through the Kadowaki-Woods relation, this implies that the effective mass}
diverges on approach to the critical field $B_c=\mu_0 H_c$ from either side of the QCP. Data for $H \perp c$ have been scaled by a factor of 11. Adapted from Ref. \cite{Gegenwart.02}.}
\label{fig:yrs_Acoeff}
\end{figure}

On the non-magnetic side, $B>B_c$, the mass enhancement diverges as $B$ approaches $B_c$ from above.
This has been 
established
by  measurements of both the $T$-linear specific-heat coefficient $\gamma$, which is proportional to the effective mass $m^*$,
and 
the $T^2$ coefficient $A$ of the resistivity, which was found to obey the Kadowaki-Woods relation \cite{Tsujii.05}. 
The divergence of $A$ is shown in 
Fig.\,\ref{fig:yrs_Acoeff}.

For $B<B_c$, the mass enhancement is also large. This is compatible with the large $C/T$ measured in the antiferromagnetic state,
although to reliably extract $\gamma$ is a challenge because of the interference 
of the large specific-heat feature at the 
magnetic transition temperature $T_N$. The mass enhancement can be more reliably extracted from the $A$ coefficient, because 
the effect of the magnetic transition at $T_N$ on the resistivity is relatively minor. 
The evolution of the $A$ coefficient 
with $B$ is consistent with the destruction of the Kondo effect as the QCP is 
approached from the nonmagnetic
side 
as well as
 the dynamical Kondo effect inside the antiferromagnetic phase.

The effect of increasing temperature on the Hall crossover can be quantified in terms of the ratio
of the crossover width to the crossover magnetic field. For $T \gtrsim 0.5$ K,
the ratio quickly increases toward unity as shown in 
Fig.\,\ref{fig:yrs_figure1}(c). This implies that, for such temperatures,
YbRh$_2$Si$_2$  
 falls in
the quantum-critical fluctuation regime already for zero magnetic field.
Thus, the
   \qs{single-particle}
    spectral weight will be 
significant at both the small
and large Fermi surfaces. In this temperature
range, 
significant spectral weight is 
thus
to be expected at the large Fermi surface.
The ARPES measurements in YbRh$_2$Si$_2$,
which have been reported for
 $T>1\mbox{K}$ \cite{Kummer.15},
are consistent with this prediction \cite{Paschen.16}.

\begin{figure}[t!]
\centering
\includegraphics[width=0.9 \linewidth]{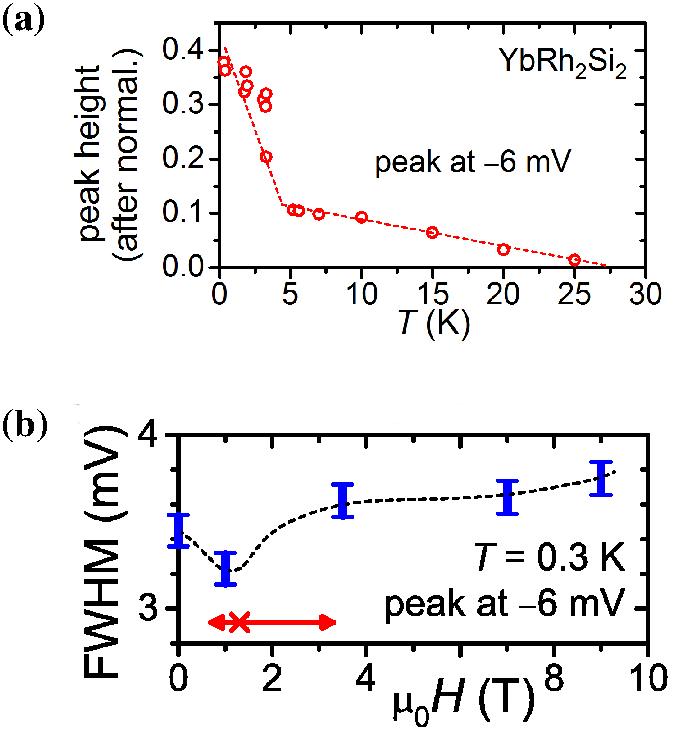}
\caption{STM spectroscopy of the lattice Kondo feature at -6meV in YbRh$_2$Si$_2$. (a) The 
temperature evolution of the peak height  of the -6meV peak. A strong increase in the peak height is 
observed below $5\mbox{K}$. 
(b) The FWHM of the -6meV peak across the critical field at the 
base temperature $T=0.3\mbox{K}$. 
Note that at this temperature all field values place the system within the quantum-critical fan.  The decrease of the peak width near $B^*=\mu_0 H^*$($T=0.3$ K) is consistent with a critical slowing down at quantum criticality.
From \cite{Seiro.18}.
}
\label{fig:yrs_stem}
\end{figure}

The temperature evolution of the single-particle excitations in YbRh$_2$Si$_2$ 
has been studied by STM measurements,
which were
first carried out down to 4.6 K
by \cite{Ernst.11}
and were recently extended down to $0.3\mbox{K}$ \cite{Seiro.18}.
The lattice Kondo effect has been identified with the feature at a particular bias, $-6\mbox{meV}$.
The initial onset of this feature takes place near $25\mbox{K}$, which corresponds to $T_{\mbox{\tiny 0}}^{\mbox{\tiny en}}$,  an estimate of $T_0$ based on the spin entropy $S$ and defined through $S(T_{\mbox{\tiny 0}}^{\mbox{\tiny en}}/2)= 0.4 R \ln 2$, where $R$ is the ideal gas constant: see Table \ref{tab:energyscales}.
At $B=0$, the measurements down to $T=0.3\mbox{K}$ show an increase in the spectral weight: see
Fig.\,\ref{fig:yrs_stem}(a).
This is compatible with the dynamical
Kondo effect at nonzero temperatures.

The STM experiments have also determined the isothermal $B$ dependence of the peak width
at the lowest measured temperature $T=0.3\mbox{K}$. It shows a minimum near $B^*$($T=0.3\mbox{K}$) 
as shown in Fig.\,\ref{fig:yrs_stem}(b). 
This observation is consistent with a critical slowing down associated with the Kondo destruction energy scale
	that was implicated by magnetotransport and thermodynamic measurements
	\cite{Paschen.04,Gegenwart.07,Friedemann.10}. As such, it represents
 the most direct evidence so far for the Kondo destruction quantum criticality based on 
 a
 single-particle measurement
 in YbRh$_2$Si$_2$.

\subsection*{Summary of Sec.\,\ref{sec:YRS}}
	We now summarize the 
	salient 
	results 
on YbRh$_2$Si$_2$ 
discussed in this section.

{\itshape High-energy features:} STM experiments for YbRh$_2$Si$_2$ at $B=0$ clearly observe the initial onset of  dynamical Kondo correlations around $T_0$,
a comparatively high temperature, as expected  for any Kondo lattice system regardless of the nature
(Kondo screened or Kondo destruction)
of its ground state. This is consistent with the observation of a
hybridization gap in the optical spectrum \cite{Kimura.06}.
As temperature is further lowered below $T_0$, 
4$f$ electron
spectral weight is expected to develop, and this has also been
clearly observed.

{\itshape Low-energy
	isotherms:}
STM experiments for YbRh$_2$Si$_2$ have been carried out as a function of magnetic field at 
$T=0.3\mbox{K}$. The Kondo lattice spectral peak shows a critical slowing-down feature at $B^*$, the 
Kondo destruction
scale 
previously determined from magnetotransport and thermodynamic measurements. As such, the STM results are consistent with 
local quantum criticality.

\section{The Cerium-based 115 Family: Photoemission versus Tunneling Spectroscopy}
\label{sec:115}

The cerium-based 115 family is comprised of compounds Ce$M$In$_5$ where $M=$ Co, Rh, or Ir. These compounds are stoichiometric and can be grown in a very clean form. All three compounds crystallize in the HoCoGa$_5$ structure type and thus possess tetragonal unit cells.
Because of their proximity to quantum criticality, they have contributed considerably to a global understanding of 
quantum-critical heavy-electron materials \cite{Park.09,Si.06}.
While CeCoIn$_5$ and CeIrIn$_5$ under ambient conditions are low-temperature superconductors, CeRhIn$_5$ is an antiferromagnet \cite{Petrovic.01,Movshovich.01}. 
In addition, substituted variants, {\itshape e.g.}, by Cd substitution on the In site or by substitution of Ce, have also been investigated; for 
 a review, see  \cite{Thompson.12}. 

\subsection{CeIrIn$_5$}
CeIrIn$_5$ at ambient pressure is a heavy-electron superconductor with a transition temperature $T_c=0.40\mbox{K}$ \cite{Petrovic.01b}. After almost two decades of study, the origin of superconductivity remains controversial, although there is growing support for a 
magnetically driven 
mechanism \cite{Chen.15}. In spite of this controversy, superconductivity in CeIrIn$_5$ has attracted recent attention because of its unusual strain tunability \cite{Bachmann.19}.

In contrast to CeCoIn$_5$ or even CeRhIn$_5$,   CeIrIn$_5$  has been comparatively less  studied by STM and ARPES. Early ARPES studies led to different conclusions concerning the formation of 4$f$-derived flat bands \cite{Fujimori.03,Fujimori.06}. 
More recently, a high-resolution ARPES study by \cite{Chen.18b}  mapped out the full bandstructure of CeIrIn$_5$. 
Interestingly, this study was able to resolve the complete fine structure of both the $4f_{5/2}^{1}$ and $4f_{7/2}^{1}$ peaks in the measured energy-distribution curves (EDCs) and momentum-distribution curves, which may be a reflection 
of the comparatively stronger $4f-c$ hybridization than in CeCoIn$_5$ \cite{Chen.18b}. 

To the best of our knowledge,  no scanning tunneling spectroscopy of CeIrIn$_5$ is available,
apart from an STM investigation that focused on 
the structural properties of CeIrIn$_5$ surfaces \cite{Ernst.10,Wirth.14}.
Our main focus in this section is therefore  on CeCoIn$_5$ and CeRhIn$_5$.

\subsection{CeCoIn$_5$}
CeCoIn$_5$ has attracted interest not only for its comparatively high superconducting transition temperature $T_c \sim 2.3\mbox{K}$ but also for an overall phenomenology that resembles that of the underdoped cuprates.

The strong interest in CeCoIn$_5$ includes early photoemission studies which, however, have led to contradictory results concerning the localized versus itinerant nature of the  4$f$ electrons \cite{Koitzsch.08,Koitzsch.09,Koitzsch.13}. Optical conductivity measurements of CeCoIn$_5$ show the existence of a hybridization gap at high energies which starts forming at comparatively high temperatures \cite{Singley.02,Burch.07} and recent  STM studies of CeCoIn$_5$ are in line with these findings \cite{Aynajian.12,Allan.13,Zhou.13}.
de Haas-van Alphen (dHvA) studies performed at low temperatures indicate that the Fermi surface of CeCoIn$_5$  includes the 4$f$ electrons and that therefore the  Fermi surface of CeCoIn$_5$ is large \cite{Settai.01,Shishido.02}. This conclusion is further corroborated by band-structure calculations that treat the 4$f$ electrons as fully itinerant \cite{Haule.10}. 

\begin{figure}[htp]
	\centering
	\includegraphics[width= \linewidth]{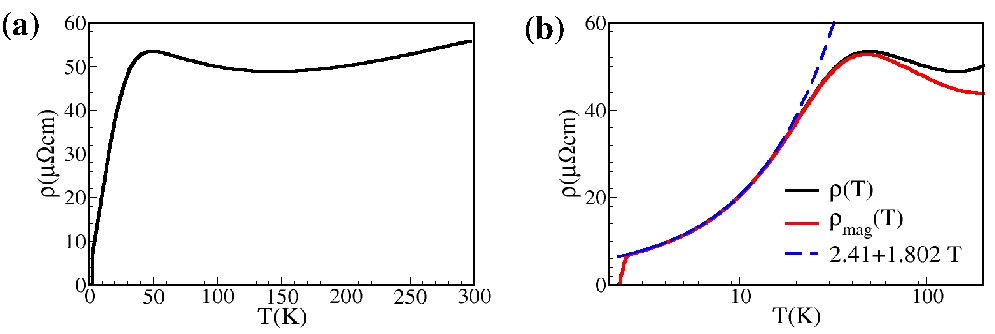}
	\caption{Temperature dependence of the resistivity $\rho$ of  CeCoIn$_5$.
		(a) $\rho(T)$ in the temperature range from $2.2$ to $300\mbox{K}$. $\rho$ has a smooth maximum around $T_{\mbox{\tiny coh}}\approx 40\mbox{K}$;
		\qs{this temperature for the resistivity maximum is commonly referred to as the coherence temperature,
		which is a manifestation of $T_0$ defined earlier.}
		(b) $\rho$ (black continuous line) and the magnetic resistivity $\rho_m$ (red continuous line) in  a semilog plot for temperatures from $2$ to $200\mbox{K}$.
		$\rho_m(T)$ is defined as the difference  between the resistivity of CeCoIn$_5$ and that of
		its nonmagnetic reference compound LaCoIn$_5$ at temperature $T$.
		The dashed   line
		represents a linear  law fit to $\rho_m(T)$ and shows that $\rho_m(T)$ is linear in $T$ from $T_c$ to approximately $20\mbox{K}$.
	}
	\label{fig:RHOCeCoIn5}
\end{figure}

CeCoIn$_5$ under ambient conditions is believed to be located close to an 
antiferromagnetic QCP
of the 
SDW
type and  can be tuned to  a quantum phase transition by applying  a magnetic field \cite{Ronning.05,Singh.07,Zaum.11}. 
The STM study by \cite{Aynajian.12} 
also reported an interesting energy-over-temperature ($\omega/T$) scaling of the local conductance of CeCoIn$_5$ which sets in  around $60\mbox{K}$. 
It is worth recalling that  STM probes the 
   \qs{single-particle}
    response while the 
    \qs{dynamical spin susceptibility}
    measures the magnetic fluctuation spectrum.
As $\omega/T$ scaling is not expected in the dynamical spin susceptibility at a QCP of the SDW type,
the observation of dynamical scaling in the local conductance suggests that the SDW nature applies, at least  at ambient conditions, only at asymptotically
low energies. 
In any case, 
the observation  of $\omega/T$ scaling does  
appear
 to be 
in line with the  linear-in-temperature behavior of the resistivity below $20\mbox{K}$  \cite{Petrovic.01} as shown in Fig.\,\ref{fig:RHOCeCoIn5}.
Further support in favor  of  such an $\omega/T$ scaling in CeCoIn$_5$ for the single-particle excitations near the $\Gamma$ point  has come from 
a recent high-resolution ARPES study  \cite{Chen.17}.

\begin{figure}[tp]
	\centering
	\includegraphics[width= \linewidth]{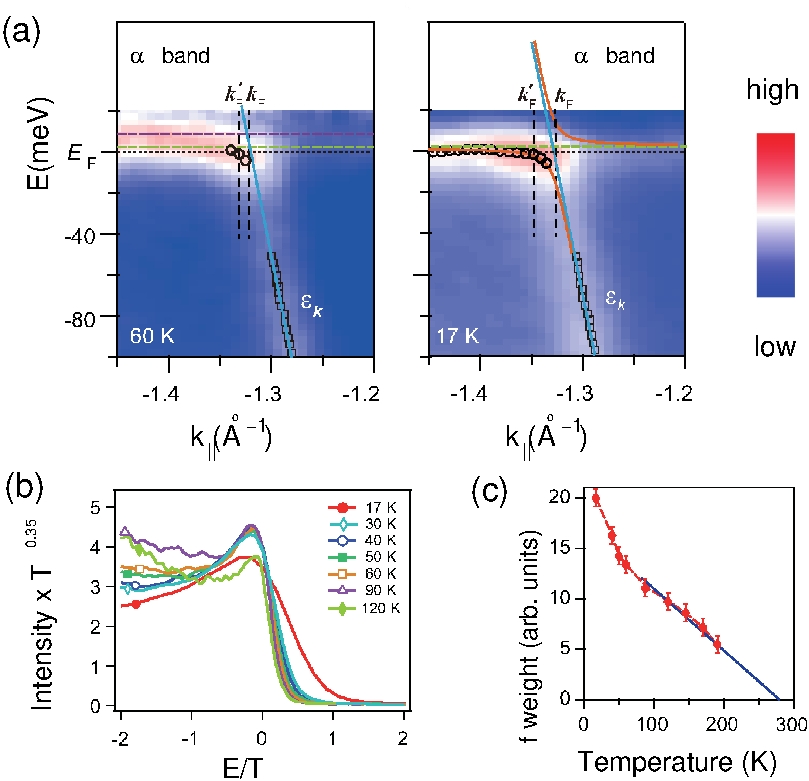}
	\caption{ARPES view 
		of the 4$f$ electron weight near the Fermi energy
		in CeCoIn$_5$. (a) Evidence for the
		initial development of
		hybridization between 4$f$ and conduction band $\alpha$ at  $T=60$ and   $T=17\mbox{K}$ after dividing the  EDCs by the Fermi-Dirac function. (b) $\omega/T$ scaling of the EDCs near the $\Gamma$ point in an intermediate temperature range and energy range around the Fermi energy $E_F$. 
		(c) Background subtracted 4$f$ electron spectral weight transfer near the $\Gamma$ point vs temperature. The EDCs have been integrated over an energy window from -40 to 2meV.
		(From \cite{Chen.17}).
	}
	\label{fig:CeCoIn5}
\end{figure}


The ARPES  study by \cite{Chen.17}  reported  the first 3D Fermi surface mapping of 
CeCoIn$_5$ and provided a measurement of the full band structure of this heavy-electron system. Because of the large temperature range of the study  from $14$ to $310\mbox{K}$, Chen et al. were able to demonstrate that the formation of the 4$f$-derived flat band sets in at temperatures far above the coherence temperature.
 This finding is significant, although not entirely unexpected.
It
demonstrates not only the slow evolution of the Kondo screening process but also the likely role of a Kondo effect on 
the 
excited crystal field levels \cite{Chen.17,Pal.19}. 
These results contrast with a temperature-independent Fermi surface in YbRh$_2$Si$_2$ that was inferred from 
the state-of-the-art
ARPES measurements in a temperature window from 1 to 100K \cite{Kummer.15}.
An earlier laser-based ARPES study of YbRh$_2$Si$_2$ reported a $T$-dependent bandstructure below 100 K \cite{Mo.12}. 
In this regard, we note that \cite{Chen.17}  suggested that ARPES at temperatures larger than 100 K may be required 
in YbRh$_2$Si$_2$ due to the large effect of the crystal field levels. 
This is consistent with $T_{\mbox{\tiny 0}}^{\mbox{\tiny hyb}}\approx$ 160K in this compound;
see Table \ref{tab:energyscales}.

High-resolution ARPES results on CeCoIn$_5$ that are largely compatible with those of Chen et al. \cite{Chen.17} have also been reported 
by Jang et al. \cite{Sooyoung-Jang.17}.
Although ARPES measurements on heavy-electron compounds have been a major experimental achievement, care has to be taken when extrapolating to the high-temperature region where the 4$f$ electrons have to be localized across the phase diagram, as previously argued.
In  
Fig.\,\ref{fig:CeCoIn5}(a)
we reproduced the  EDCs from \cite{Chen.17} for the $\alpha$ band, one of three bands that are part of  the high-temperature Fermi surface, 
in the vicinity of its Fermi crossing  for both $T=60$ and  $T=17\mbox{K}$. 
The data have been  divided
by
the Fermi-Dirac function to access the region (slightly) above the Fermi energy. The dashed lines in 
Fig.\,\ref{fig:CeCoIn5}(a)  indicate the positions of  maxima of the main and the first excited \sk{crystal electric field-related} Kondo resonancelike features, both of which are taken to be dispersionless. Here $k_F$ is the Fermi momentum of the conduction electrons without 4$f$ participation, {\itshape i.e.}, at high temperatures.

In Fig.\,\ref{fig:CeCoIn5}(c), the building up of spectral weight near the Fermi energy is shown as a function of temperature. This is calculated by integrating the EDCs near the Fermi energy, {\itshape i.e.}, from  
-40 to 2meV, and after subtracting a flat, temperature-independent overall background. It is worth recalling that the majority of the Kondo resonancelike features of a cerium-based system is located above the Fermi energy, a region which is, especially at low $T$,  inaccessible to ARPES.

It is instructive to analyze the high-resolution ARPES data of Chen et al. for the temperature-dependent band structure of CeCoIn$_5$ in light of the expectation that the Fermi surface of this compound should contain the 4$f$ electrons at sufficiently low temperatures.
In other words, in terms of Fig.\,\ref{fig:lqcp}, CeCoIn$_5$ is located on the 
$\delta>\delta_c$
side of the 
$E^*_{\rm  loc}$
line. Note, however, that 
Fig.\,\ref{fig:lqcp} presents 
one type of
specific cut through
the  global heavy-electron phase 
diagram \cite{Si.06}
\qs{; its variant, with $E^*_{\rm  loc}$ at $\delta_c$ being small but nonzero, is believed to describe CeCoIn$_5$.}
  
The continuous red line in the right-hand panel of 
Fig.\,\ref{fig:CeCoIn5}(a)
 is a fit of the data to the mean field expression for the single level, single band Anderson lattice model. 
The circles in 
Fig.\,\ref{fig:CeCoIn5}(a) are obtained from the maximum in the EDCs and interpreted as the dispersion of the quasiparticle band. This leads to the value of $k_F'$, where $k_F'$
is the {\itshape projected} zero-temperature Fermi momentum.
Such a fit should not be taken too literally.
As previously mentioned,  mean field approaches may in principle  be suitable to address the conduction bands at comparatively high energies and temperatures or the low-energy behavior on either the small  or the large Fermi  volume side  in a limited energy range.
They do, however, generically fail to describe the crossover from the high- to the low-energy or temperature behavior.
In addition, there is the general difficulty of constructing the correct mean field theory. The effective model for a  system like CeCoIn$_5$ should not be the single level, single band Anderson lattice model. Nonetheless, the mean field construction provides an estimate for  the change in the Fermi wave vector from its high-temperature value $k_F$ to $k_F'$.
If $k_F'=k_F$, the 4$f$ electrons remain localized and do not contribute to the Fermi volume.
As discussed  and also briefly mentioned in  \cite{Chen.18}, if the Fermi surface of CeCoIn$_5$ expands from $k_F$ to $k_F'$ as the zero-temperature
\qs{limit}
 is approached,
the bandstructure in the vicinity of $E_F$ should resemble that sketched in 
Fig.\,\ref{fig:sketch}(a)
and the spectral weight close to $k_F$ needs to vanish
 as  $T\rightarrow 0$ so that the incoherent spectral weight at the Fermi energy is gapped out. The detection of such a, possibly very small, gap is challenging in view of the limited energy resolution and $k_z$ broadening effects of ARPES experiments  as discussed in Sec.\ \ref{sec:ARPES-STM}.
Note that, although it is expected that $k_F'\neq k_F$ in CeCoIn$_5$, results shown in
Fig.\,\ref{fig:CeCoIn5}(a)
are indicative of a  spectral weight increase near and at $k_F$  as the temperature is lowered from $T=60$ to $T=17\mbox{K}$. 
\qs{Most likely, this is more than just a reflection of the limited energy resolution of the measurement, and instead
indicates that the single-particle excitations are not of the Fermi liquid form}
depicted in  Fig.\,\ref{fig:sketch}(a).
This is also corroborated by the strange metal behavior, encoded  in an approximately linear-in-temperature dependence of the resistivity over a wide temperature window  above the superconducting transition temperature ($T_c \sim 2.3\mbox{K}$) \cite{Petrovic.01}. 
In Fig. \ref{fig:RHOCeCoIn5}, the temperature dependence of the resistivity $\rho$ of CeCoIn$_5$ is shown together with the magnetic resistivity, {\itshape i.e.}, the difference between the resistivities of CeCoIn$_5$ and its nonmagnetic reference compound LaCoIn$_5$.

\begin{figure}[b!]
\centering
\includegraphics[width=\linewidth]{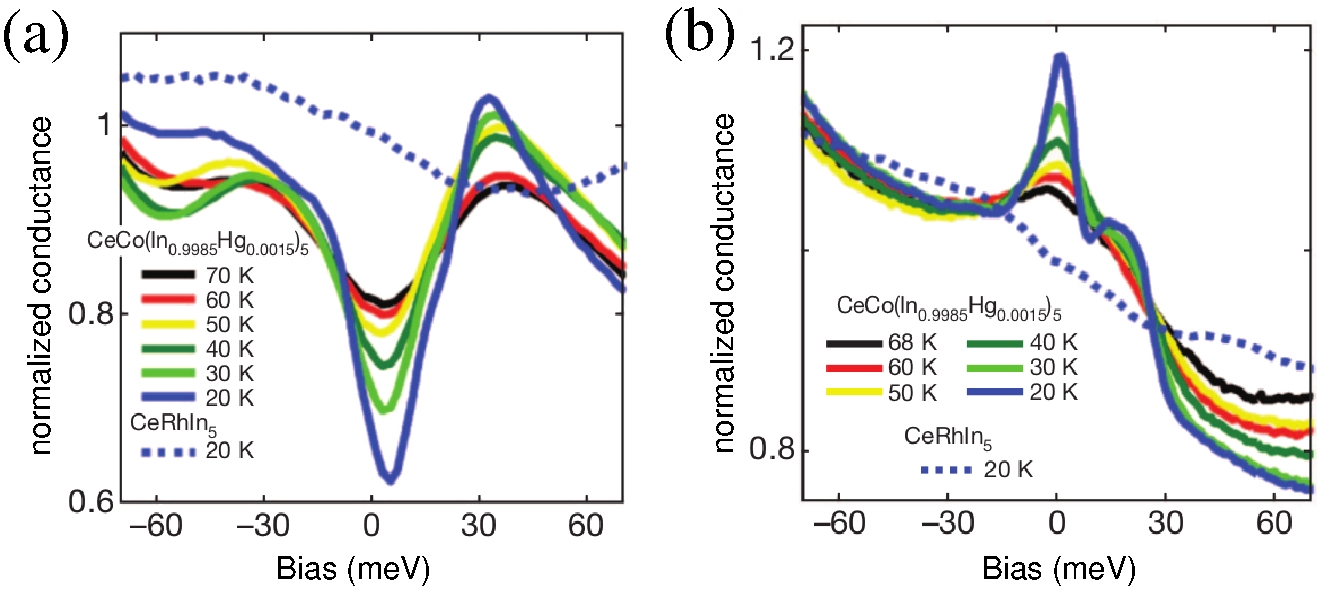}
\caption{Tunneling spectroscopy of CeCoIn$_5$ and CeRhIn$_5$: local conductance vs applied bias voltage for different temperatures  on (a) Ce-terminated surfaces  and (b) Co-(respectively, Rh-) terminated surfaces. The peak-dip-peak structure in conductance of CeCoIn$_5$ (a) is typical of a hybridization gap that is not obvious in CeRhIn$_5$, even at the lowest temperature. From \cite{Aynajian.12}.}
\label{fig:STM-115}
\end{figure}

The ARPES study of \cite{Chen.17} also
indicated
 the presence of $\omega/T$ scaling in the  EDCs near the $\Gamma$ point in an intermediate temperature range. This is reproduced in Fig.\,\ref{fig:CeCoIn5}(c). Already at around $90\mbox{K}$, the EDCs multiplied by $T^{x_{\text \tiny EDC}}$ (with $x_{\text \tiny EDC}\approx 0.36$) collapse on a function depending only on $\omega/T$. This, however, should not be  interpreted as reflecting an $\omega/T$ scaling of all single-particle excitations, which would imply a strict linear-in-$T$ behavior of the resistivity. Indeed,  this scaling seems to be confined to the vicinity of the $\Gamma$ point and is absent in the angle-integrated EDCs.
Moreover, this peculiar scaling exists only in an intermediate $T$ range and fails below $20\mbox{K}$, as shown in  
Fig.\,\ref{fig:CeCoIn5}(c). 
 This conclusion appears to be compatible with the findings reported in \cite{Aynajian.12}, taking into account that tunneling into states with small lattice momenta is favored over tunneling into large-momentum states \cite{Tersoff.85,da-Silva-Neto.13,Huang.15}.
This demonstrates that ARPES and STM indeed provide information on the single-particle Green's function that can be directly compared to each other. It is, however, noteworthy that the temperature exponents accompanying this $\omega/T$ scaling in the intermediate temperature range from $20$ to around $70\mbox{K}$ differ somewhat depending on the measurement technique. While the STM-derived exponent is $x_{\text \tiny STM}\approx 0.53$, the best fit of the ARPES data was obtained for $x_{\text \tiny EDC}\approx 0.36$. The difference between the ARPES and STM results is most likely due to the dependence of the STM current on the degree of tunneling into 4$f$ and $c$ electron states. This dependence is encoded in the Fano parameter.

STM studies on CeCoIn$_5$  (and to a much lesser extent on CeRhIn$_5$ and CeIrIn$_5$) have been performed by several groups \cite{Aynajian.12,Allan.13,Zhou.13,Aynajian.14,Haze.18,Ernst.10}.
In 
Figs.\,\ref{fig:STM-115}(a)
and (b), results are shown for the local tunneling conductance of CeCoIn$_5$ very lightly doped with mercury (Hg) as well as CeRhIn$_5$  at different temperatures and on  two different surfaces \cite{Aynajian.12}. 
The Hg-doping induced disorder in  CeCoIn$_5$  generates impurity scattering at the dopant sites which in turn can be systematically used to obtain lattice momentum-resolved information of the local DOS through  QPI \cite{Derry.15}. 
This use of QPI to extract the band structure near $E_F$ in the low-temperature limit, however, also has  potential shortcomings  that were already alluded to in Sec.\ \ref{sec:ARPES-STM}.

\begin{figure}[b!]
\centering
\includegraphics[width= \linewidth]{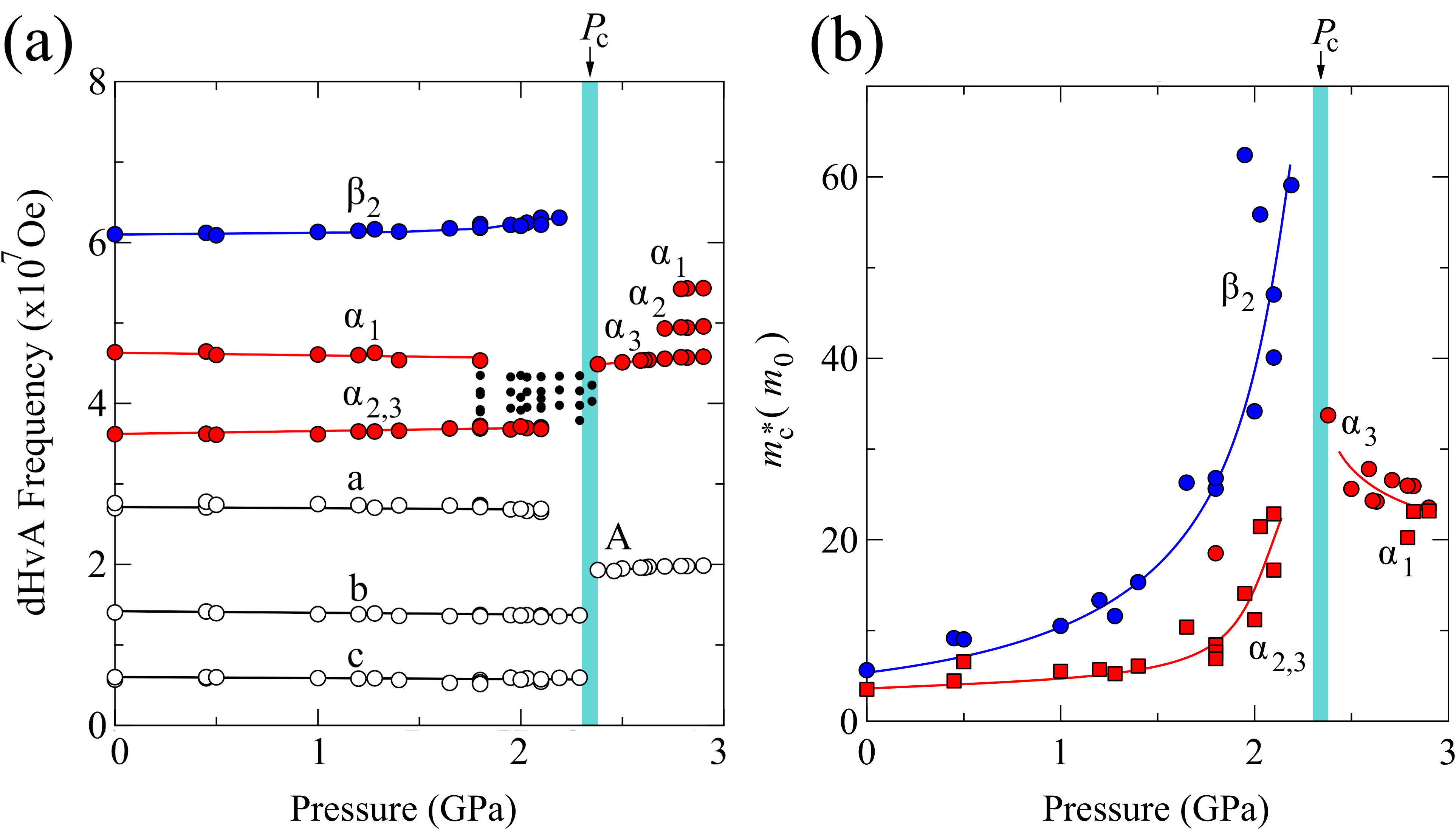}
\caption{de Haas-van Alphen  measurements  on CeRhIn$_5$. (a) Jump of the dHvA frequencies at $p_c$ indicating a reconstruction of the Fermi surface as the QCP is crossed. (b) Diverging  effective mass  upon approaching $p_c$ from above and below. Adapted from \cite{Shishido.05}}.
\label{fig:dHvASTM}
\end{figure}

\subsection{CeRhIn$_5$}
 CeRhIn$_5$ is an antiferromagnet with a N\'{e}el temperature of $T_N=3.8\mbox{K}$ at ambient pressure and has predominantly localized moments \cite{Hegger.00}. Under pressure,   $T_N$ can be suppressed to zero, thus tuning the system to a QCP at a critical pressure $p_c$. de Haas-van Alphen studies of CeRhIn$_5$ across the QCP display a clear jump of the dHvA frequencies at $p_c$, see Fig.\,\ref{fig:dHvASTM}(a), which  implies that the Fermi surface changes discontinuously at the QCP \cite{Shishido.05}. This compound therefore likely hosts a 
Kondo destruction
QCP at $\delta_c =p_c$ ($\delta$ was defined in Sec.\ \ref{sec:QCP}). This conclusion is further corroborated by an effective mass that tends to diverge
on approach to  $p_c$; see  Fig.\,\ref{fig:dHvASTM}(b). The latter  reflects the vanishing of the wave-function renormalization factor $z$, depicted in 
Fig.\,\ref{fig:lqcp}(b), as the QCP is reached from either above or below  $p_c$.
In addition, transport measurements provide evidence for the Kondo destruction QCP \cite{Park.06,Park.08}. These low-energy quantum-critical features are accompanied by experiments measuring high-energy properties. The optical conductivity of CeRhIn$_5$ was reported in \cite{Mena.05} and  shows the formation of a weak hybridization gap at high frequencies as temperature is lowered below the crossover scale $T_0$. 

Despite  evidence for the existence of a  QCP
 featuring critical reconstruction of the Fermi surface in CeRhIn$_5$ under pressure, APRES and STM investigations of this compound  are comparatively rare. This is largely due to  difficulties in preparing a suitable surface and to the present impossibility of making these measurements under applied pressure.
Early nonresonant ARPES investigations of  CeRhIn$_5$ reported  that the 4$f$ electrons in this compound are predominantly itinerant \cite{Moorea.02}, whereas  a second nonresonant ARPES study argued that the 4$f$ electrons  are nearly localized \cite{Fujimori.03}.

\begin{figure}[b!]
\centering 
\includegraphics[width= \linewidth]{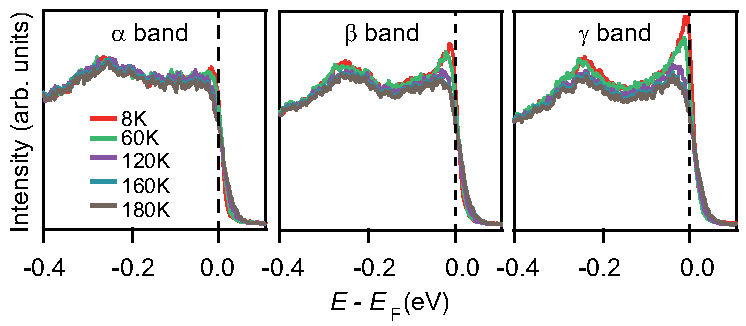}
\caption{EDCs of CeRhIn$_5$ vs temperature. The energy-distribution curves show the evolution of spectral weight with temperature  near the Fermi energy $E_F$ for the three bands that cross $E_F$, labeled  $\alpha$, $\beta$, and $\gamma$. Data were taken  along the $\Gamma$M direction at $k_{\parallel}= -0.57\AA^{-1}$ ($\alpha$ band), $k_{\parallel}= -0.3\AA^{-1}$ ($\beta$ band), and $k_{\parallel}= -0.124\AA^{-1}$ ($\gamma$ band)  and with an uncertainty of $\delta k_{\parallel} \sim 0.03 \AA^{-1}$ for each of the three $k_{\parallel}$ values. From \cite{Chen.18}.}
\label{fig:bandependence}
\end{figure}

Scanning tunneling spectroscopy data on Ce- and  Rh-terminated surfaces of CeRhIn$_5$  show no  clearly
discernible
Fano resonances, at least  at around $20\mbox{K}$ \cite{Aynajian.12}, see 
Fig.\,\ref{fig:STM-115}(a) 
and (b).
Interestingly, these results 
seem
 incompatible with high-resolution resonant ARPES data  which point to the
development of the 4$f$-electron spectral weight
 near the Fermi energy, although the weight transfer is much weaker than in CeCoIn$_5$;
see Fig.\,\ref{fig:bandependence} \cite{Chen.18}. The spectral weight transfer depicted in 
Fig.\,\ref{fig:bandependence}
for the three bands crossing the Fermi surface also shows that, in the temperature range studied, 
spectral weight transfer occurs mainly near the $\gamma$ band crossing. 
The difference between the ARPES measurements of \cite{Chen.18} 
and the STM investigation of  \cite{Aynajian.12}  
is likely due to  the increased surface sensitivity of STM. 
One possibility is that the Kondo temperature at the surface is reduced due to the reduced hybridization;
a second is that the cleaving process to obtain suitable surfaces appears to be  more problematic for CeRhIn$_5$ and CeIrIn$_5$ than
 for CeCoIn$_5$. In fact,  recent STM 
results \cite{Haze.19}
 on epitaxially grown CeRhIn$_5$ with well-defined surfaces are very much in line with the  
 ARPES measurements of \cite{Chen.18}.
As in the case of
  the STM images of YbRh$_2$Si$_2$
[see Fig.\ \ref{fig:yrs_stem}(a)],
 these data are consistent with the dynamical Kondo effect taking place near the small Fermi surface.

\subsection{Further considerations}

We now turn to several additional points that cut across specific Ce-115 families.
 First, the connection between the different Ce-115 families deserves further studies.
As already discussed, isothermal dHvA measurements in CeRhIn$_5$ provide evidence for 
a sudden Fermi surface reconstruction at $p_c$. 
Intriguingly,
 frequencies of dominant $\alpha$ orbits at $p>p_c$ for CeRhIn$_5$ are very similar to those found 
 for the large Fermi surface of CeCoIn$_5$ at atmospheric pressure \cite{Shishido.05}.
No Fermi-surface reconstruction was found in CeCoIn$_5$. 
It is possible that a sudden Fermi-surface reconstruction can still be found 
in CeCoIn$_5$ under a new tuning parameter, such as negative pressure.
But it may also be that such an effect simply does not exist in 
CeCoIn$_5$, reflecting its inherent difference from CeRhIn$_5$.
For example, 
the $4f-c$ hybridization is much larger in CeCoIn$_5$ than in CeRhIn$_5$, 
as evident in their STM spectra; see Fig. \ref{fig:STM-115}. As argued recently, this difference in overall hybridization 
can be traced to an anisotropic spatial extent of their $4f$ orbitals that is set by details of the 
crystal electric field wavefunction \cite{Willers.15,Sundermann.19}.

Second, an alternative explanation for the jump of the dHvA measurements
across $p_c$ in CeRhIn$_5$ was
proposed by \cite{Watanabe.10}.  It was suggested that the $4f$-valence fluctuations
lead to a rapid valence change near $p_c$ and a strongly first order 
antiferromagnetic transition.
The latter implies a large jump of the order parameter and, thus 
a large reconstruction of the Fermi surface.
So far, however,  all experimental evidence points to a continuous transition at $p_c$.
In addition, canonical valence-fluctuating systems such as CeSn$_3$ (CePd$_3$) have 
specific-heat coefficients of $53$ ($37$) mJ/(mol K$^2$), 
\qs{and effective}
Kondo temperatures 
of $770$ ($1120$) K \cite{Lawrence.81}.
In those cases, the $4f$-occupancy
$n_f$ will be far from $1$ or $0$ and, consequently, the entropy in the valence-fluctuation sector, 
which one can estimate by $R \left [ n_f \ln n_f^{-1} + (1-n_f) \ln (1-n_f)^{-1} \right ]$,
will be a sizable fraction of $R \ln 2$. By contrast, in the quantum-critical regime of CeRhIn$_5$,
 the specific-heat coefficient is very large [$\gamma \approx 1.25$ J/(mol K$^2$)] \cite{Park.09},
 implying that $n_f$ is exceedingly close to $1$. Thus, the valence-fluctuation sector
 will have a small entropy compared to the nearly $R \ln 2$ entropy in the spin sector and can 
 hardly be the main driver of the critical fluctuations.
 In other words, the quantum criticality should primarily be driven by physics 
 of the Kondo limit \cite{Park.06,Park.08}. 
Similar arguments apply to CeCoIn$_5$,
CeIrIn$_5$, YbRh$_2$Si$_2$, and CeCu$_{5.9}$Au$_{0.1}$.

\subsection{Summary of Sec. \,\ref{sec:115}}

We close this section by summarizing  the status of ARPES and STM investigations in cerium-based 115 systems as discussed in this section.

By and large, the existing  STM and ARPES results on the  cerium-based
115 family are  consistent with each other, given the requirements of surface quality and the associated difficulties.  
The recent high-resolution ARPES investigation of
these 115 
materials
also shows that none of the three compounds follows the low-temperature band-structure  expectations of
\qs{a fully-coherent heavy-electron Fermi liquid,}
encoded in 
Fig.\,\ref{fig:sketch}(a).
This is in line with other measurements, in particular transport measurements, which suggest that none is in a Fermi liquid regime in the range where the ARPES measurements were made. 
Further, the limited energy resolution of state-of-the-art ARPES is still posing a major challenge in the heavy-electron materials class in which the associated energy scales are typically very small.

{\itshape High-energy features:} Existing  ARPES and STM investigations of the 115 members   show the
initial onset of dynamical Kondo correlations
  around 
 the
 $T_0$
 temperature scale [Fig.\, \ref{fig:lqcp}(a)]
 and the concomitant 
 onset
  of  
  hybridization-gap
  formation. This is in line with optical conductivity measurements on these compounds \cite{Chen-Wang.16}. 
  Comparing ARPES and STM data for the same compound gives complementary results that are compatible
  with each other, and provide evidence for the existence of the hybridization-gap onset scale $T_0$.

{\itshape Low-energy features:}  Neither in CeCoIn$_5$ nor in CeRhIn$_5$ has ARPES been able to confirm unambiguously the existence of either $k_F^L$ or $k_F^S$. While this may not be  surprising due to the limited  energy and momentum resolution currently available to ARPES, 
this finding is also compatible with the absence of Fermi liquid signatures
 in the investigated temperature range in these compounds;
 in this range, Fermi liquid signatures
 are absent as well in transport and thermodynamic properties.
Isothermal measurements of dHvA have shown a sudden reconstruction of the Fermi surface 
across the pressure-induced  QCP in CeRhIn$_5$, which provides strong evidence 
for a Kondo destruction QCP.\\[2ex]
\phantom{.}
\vskip 20pt
%
%
%
%
%
\onecolumngrid
\renewcommand{\arraystretch}{1.5}
\begin{table*}[ht!]
	\hspace*{-1.5cm}
	\begin{tabular}{|l |l |l |l |l|}
		\hline
		\rowcolor{gray!50} 
		& $T_{\mbox{\tiny  FL}}$(K) & $T_{\mbox{\tiny 0}}^{\mbox{\tiny hyb}}$(K) &$T_{\mbox{\tiny 0}}^{\mbox{\tiny en}}$(K) & Reference\\ 
		\midrule 
		YbRh$_2$Si$_2$ & 0.07  & $\sim$ 160  & $\approx 24$ & \cite{Trovarelli.00,Kimura.06,Gegenwart.06} \\
		\midrule  
		YbRh$_2$Si$_2$  & \multirow{2}{*}{$<0.008$ (LMT)}   & \multirow{2}{*}{$\sim$ 160}  &\multirow{2}{*}{$\approx 24$} & \multirow{2}{*}{\cite{Taupin.15,Kimura.06,Gegenwart.06} }\\
		~(B=B$_c$, B$\parallel$ c)& & & & \\ 
		\midrule  
		YbRh$_2$Si$_2$  & \multirow{2}{*}{0.135}  & \multirow{2}{*}{$\sim$ 160}  &\multirow{2}{*}{$\approx 24$ } & \multirow{2}{*}{\cite{Gegenwart.02,Kimura.06,Gegenwart.06} }\\
		~(B=2T, B$\parallel$ c)& & & & \\ 
		\midrule 
		CeCoIn$_5$ & \multirow{1}{*}{0.14}  & \multirow{1}{*}{$\gtrsim$ 100}  & \multirow{1}{*}{$\approx 25$ }& \multirow{1}{*}{\cite{Paglione.07,Mena.05,Petrovic.01}} \\
		&(B=6T) & & & \\ 
		\midrule 
		CeRhIn$_5$ &   \multirow{1}{*}{$<$ 0.15} & \multirow{1}{*}{$\gtrsim 60^{(\dagger)}$}  & \multirow{1}{*}{$\approx$ 10} & \multirow{1}{*}{\cite{Park.08,Chen.18,Park.09}}\\
		&(p$_c=2.35$GPa, $\mu_0$H=10T) & & & \\ 
		\midrule 
		CeCu$_6$ & 0.2  &  $\gtrsim$ 40  & $\approx$ 4 &\cite{Amato.87,Marabelli.90,Fischer.87}\\ 
		\midrule  
		CeCu$_{6-x}$Au$_{x}$\,($x_c =0.1$)
		&  $<0.02$ (LMT) &  $\gtrsim$ 40  &$\approx$ 4 &\cite{Loehneysen.94,Marabelli.90,Fischer.87}\\ 
		\hline
	\end{tabular} 
	\caption{Characteristic high- and low-temperature scales for several heavy-electron compounds located in the vicinity of quantum criticality. 
		Here
		$T_{\mbox{\tiny  FL}}$ is a temperature scale below which
		the
		Landau Fermi liquid $T^2$ resistivity
		is observed. 
		$T_{\mbox{\tiny 0}}^{\mbox{\tiny hyb}}$ 
		is a ''high temperature'' estimate for the onset of the hybridization gap 
		and is estimated from the optical conductivity $\sigma(\omega,T)$ 
		(Figs.\,\ref{fig:optcond} and \ref{fig:opticalconductivity})
		with the exception of CeRhIn$_5$, 
		where existing $\sigma(\omega,T)$ data 
		indicate only that $8\mbox{K}<T_{\mbox{\tiny 0}}^{\mbox{\tiny hyb}}<300\mbox{K}$ \cite{Mena.05}. 
		$T_{\mbox{\tiny 0}}^{\mbox{\tiny en}}$ is a ''low-temperature''
		estimate of $T_0$ based on the spin entropy $S$, 
		using a procedure for the single-impurity Kondo model with constant conduction electron density 
		of states (for which $T_{\mbox{\tiny 0}}$=T$_{\mbox{\tiny 0}}^{\mbox{\tiny en}}$=$T_K^0$): 
		$S(T_{\mbox{\tiny 0}}^{\mbox{\tiny en}}/2)= 0.4 R \ln 2\approx 0.277 R$, 
		where $R=8314.5$ mJ/(mol K) is the ideal gas constant.
		LMT designates the lowest 
		measured
		temperature for the electrical resistivity
		$\rho$.\\ 
		For YbRh$_2$Si$_2$ at the critical field, $T_{\mbox{\tiny FL}}$ has been estimated from $\rho(T)$ and 
		using 
		the result
		that $\rho(T)\sim T$ down to 
		the LMT of 8mK \cite{Taupin.15},
		making the listed value to be an upper bound.
		The hybridization gap onset in 
		\qs{$\sigma(\omega)$} 
		is assumed to be the same for $0\leq B\leq 2$T. 
		Similarly,   changes of $T_{\mbox{\tiny 0}}^{\mbox{\tiny en}}$
		are assumed to be small
		for fields $0\leq B\leq 2$T, 
		where  the specific heat at around $20\mbox{K}$ is only weakly field dependent for  $B\leq 2T$ \cite{Gegenwart.06}.\\
		For CeRhIn$_5$, the QCP is located at p$_c=2.35$GPa and 
		H$_c$ with $\mu_0$H$_c \lessapprox$ 10T \cite{Park.08}.\\
		$^{(\dagger)}$This value for $T_{\mbox{\tiny 0}}^{\mbox{\tiny hyb}}$ has been estimated from the ARPES data of \cite{Chen.18} for ambient conditions; see also Fig.\,\ref{fig:bandependence}. \\
		For   CeCu$_{6-x}$Au$_{0.1}$, $\rho(T)$ 
		is linear in $T$ down to the LMT \sk{of} 20mK; 
		hence, the listed value is also an upper bound. The estimate of
		$T_{\mbox{\tiny 0}}^{\mbox{\tiny hyb}}$ in CeCu$_{6-x}$Au$_{0.1}$ is supported by the specific heat data of \cite{Loehneysen.94}.\\
		The references in this table are arranged such that in each row the first reference provides T$_{\mbox{\tiny  FL}}$, the second contains estimates for   $T_{\mbox{\tiny 0}}^{\mbox{\tiny hyb}}$ and the third provides results on the low-temperature (spin) entropy.
	}
	\label{tab:energyscales}
\end{table*}
\twocolumngrid

\phantom{.}
\vskip 1pt
\section{Progress, Challenges, and Prospects}
\label{sec:PCO}

\subsection{High-energy Kondo features}

We have stressed that the initial onset of dynamical Kondo correlations or hybridization is expected, at the $T_0$ scale of 
Fig.\,\ref{fig:lqcp}, 
for all heavy-electron systems regardless of the nature of their ground states.
\begin{figure}[h!]
\centering
\includegraphics[width=0.62 \linewidth]{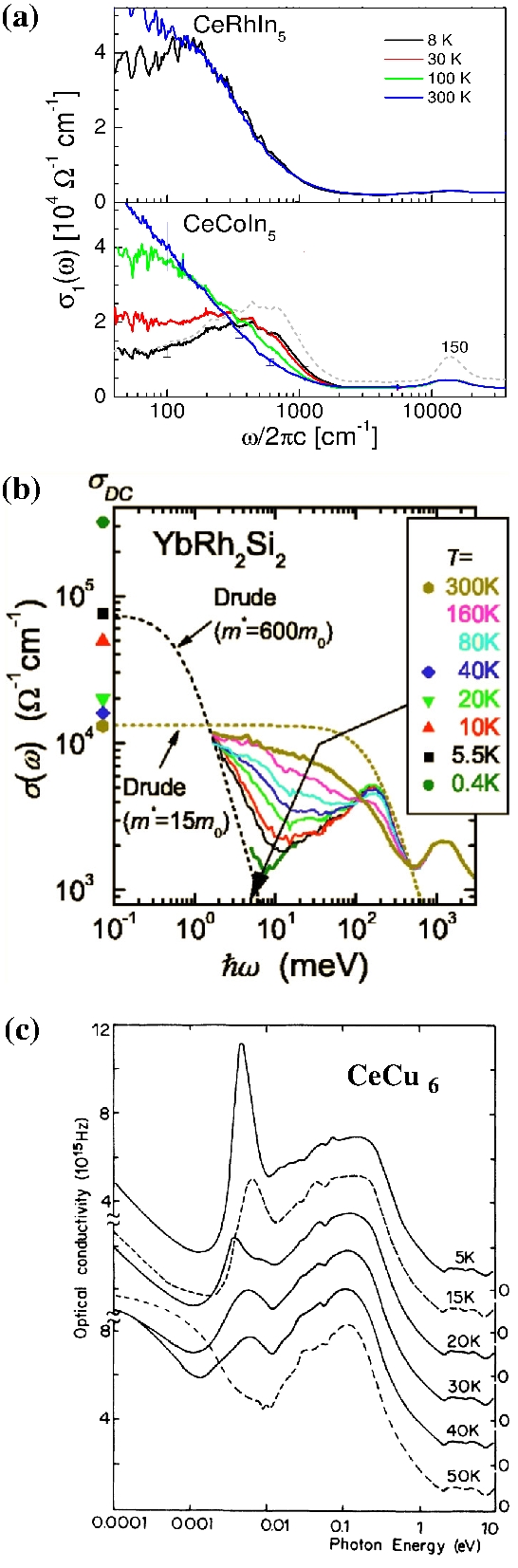}
\caption{Optical conductivity $\sigma(\omega,T)$ and evolution of the hybridization gap. (a) Although the hybridization gap in CeRhIn$_5$ (top) is overall less pronounced than that in the optical conductivity of CeCoIn$_5$ (bottom),  the overall features for both compounds are in accordance with general expectations
 (see Fig.\,\ref{fig:optcond}):
 at the highest measured $T$ a broad Drude peak exists out of which a hybridization gap develops below $\sqrt{T_0 D}$  as $T$ is lowered. Adapted from \cite{Mena.05}.
(b) The hybridization gap in  YbRh$_2$Si$_2$ evolves over a large $T$ region, starting well above $100\mbox{K}$. As the data are taken at zero external field, the system is located on the $\delta<\delta_c$ side (see Sec. \ref{sec:QCP}) and a Drude peak is therefore expected in $\sigma(\omega,T)$ at small $\omega$ and sufficiently low $T$. From \cite{Kimura.06}. 
(c) In CeCu$_6$ the optical conductivity develops a hybridization gap at around $\hbar \omega\approx 1\mbox{meV}$ below $50\mbox{K}$ which is  flanked toward higher energies by a pronounced peak.
From \cite{Marabelli.90}.
}
\label{fig:opticalconductivity}
\end{figure}
This scale is evident in YbRh$_2$Si$_2$ by STM and optical conductivity \sk{[Fig. \ref{fig:opticalconductivity}(b)]}. Similarly, the formation of a
hybridization gap was manifested in CeRhIn$_5$ by optical conductivity measurements \cite{Mena.05}, \sk{reproduced in Fig.\ \ref{fig:opticalconductivity}(a),} and, recently, by 
STM measurements \cite{Haze.19}. 
Also for CeCu$_6$, which is 
near
a QCP that is accessed by introducing 
Au substitution for Cu, a hybridization gap has been observed \sk{in the optical conductivity, see Fig.\ \ref{fig:opticalconductivity}(c)} \cite{Marabelli.90}.
This captures the high-energy $T_0$ scale for the onset of hybridization-gap formation [Fig.\,\ref{fig:lqcp}(a)]
and indeed evolves smoothly across the critical substitution $x_c=0.1$ based on
photoemission measurements \cite{Klein.08}. The 
$T_0$ scale is also evidenced by recent time-resolved 
measurements in the critical substitution range \cite{Wetli.18,Pal.19}.
Here, a terahertz irradiation pumps the system and disturbs the correlations between the local moments and
conduction electrons. We can expect the underlying Kondo coupling to produce an initial echo at a time
corresponding to $\hbar/(k_B T_0)$. Such a finite timescale is indeed observed both away from and at the QCP. 
Note that the Fermi liquid scale of CeCu$_6$ is $0.2\mbox{K}$ (see Table \ref{tab:energyscales}), which is not accessible by current 
experiments done at temperatures above $1.5\mbox{K}$. Nonetheless, it is conceivable that future experiments 
may probe not only the echo effect at $\hbar/(k_B T_0)$, 
\qs{but also the response in}
 the scaling time regime much beyond $\hbar/(k_B T_0)$.

Table \ref{tab:energyscales} compiles  high-temperature and Fermi liquid energy scales 
of the heavy-electron compounds discussed in this Colloquium.
This table lists both 
$T_{\mbox{\tiny 0}}^{\mbox{\tiny hyb}}$, the initial onset of the hybridization gap,
and $T_{\mbox{\tiny 0}}^{\mbox{\tiny en}}$,
based on the spin entropy $S$.
These two high-energy scales can differ by 
as much as an order of magnitude, which is not too surprising given that 
the crossover of Kondo lattice systems from the high-temperature incoherent regime toward the 
low-temperature 
\qs{coherent, quantum critical, or ordered}
 regime is rather broad.
 \qs{This crossover}
  can be made even 
broader 
when the excited crystal field levels are involved. In practice, we propose to use
\begin{eqnarray}
T_0 = 
\sqrt{T_{\mbox{\tiny 0}}^{\mbox{\tiny hyb}}
~T_{\mbox{\tiny 0}}^{\mbox{\tiny en}}}
\label{eq:T0-geomean}
\end{eqnarray}
as a measure of the crossover Kondo scale.
Defined in this way, we can infer from Table \ref{tab:energyscales} that 
$T_0$ is $\sim 62\mbox{K}$ in YbRh$_2$Si$_2$, $\sim 50\mbox{K}$ in CeCoIn$_5$, 
$\gtrsim 25\mbox{K}$  in CeRhIn$_5$,
and $\gtrsim 13\mbox{K}$  in CeCu$_6$.

\subsection{Isothermal evolution at low temperatures}

We  discussed in Sec.\ \ref{sec:QCP} that,
 to assess the nature of quantum criticality (Kondo destruction versus SDW), 
the
 isothermal evolution of  quasiparticle spectral weight at low temperatures 
 is
 particularly informative.
  In YbRh$_2$Si$_2$, this was done through STM measurements as a function of magnetic field at $T=0.3$ K,
and the results \cite{Seiro.18}
 support the Kondo-destruction scale that had been inferred from magnetotransport and 
  thermodynamic measurements \cite{Paschen.04,Gegenwart.07,Friedemann.10}.
Further STM  measurements at lower
temperatures will clearly be instructive.
Whether related STM studies can be carried out in 115 systems is at the present time unclear, because the QCP is realized at a
relatively large pressure (CeRhIn$_5$) or possibly at negative pressure (CeCoIn$_5$ and CeIrIn$_5$) \cite{Sidorov.02,Pham.06}. In these latter two cases, applying uniaxial  tension might open the possibility of both ARPES and STM studies in a regime that would access their 
respective QCP. Similar
isothermal studies by ARPES appear to be difficult, due to the low temperature that is needed, and also because ARPES cannot be performed in the presence of a magnetic field.

\subsection{Outlook}

As previously discussed  (see Sec.\ \ref{sec:ARPES-STM}), 
STM is a real
space probe and thus generally lacks  momentum resolution.
It is, however, possible to extract information on the band structure near the Fermi energy using Friedel oscillations that occur near defects \cite{Peterson.98,Peterson.00}.
Since STM is a surface probe,  
QPI
 provides only a projected band structure. Furthermore, the standard approach which is based on Born scattering 
is known to be insufficient in many cases \cite{Toldin.13}.  This 
limitation
notwithstanding, it 
will
 be instructive to obtain band structure information through Fourier transform STM on either side of the QCP 
  to interpret QPI spectra in the quantum-critical fan of the QCP.

Critical Kondo 
destruction
 is accompanied by a particular 
kind of $\omega/T$ scaling. Recently, this type of scaling 
was demonstrated for the optical conductivity of YbRh$_2$Si$_2$  thin films grown by molecular beam epitaxy, 
studied by time-domain THz-transmisson  spectroscopy \cite{Prochaska.18}.  
This result would be surprising from the perspective of an SDW QCP,
where only the spin dynamics is expected to be critical.
However, it is in line with critical Kondo destruction \cite{Prochaska.18}.
Because the magnetic quantum phase transition is accompanied by the
transition from a phase with asymptotically decoupled local-moment and conduction electron degrees of freedom 
to one in which the entangling of the two turns the $4f$ local moments into composite quasiparticles,
it is natural that both the single-particle and charge dynamics are critical. 
Indeed,
calculations at the Kondo destruction QCP in
various large-$N$ limits  \cite{Zhu.04,Kirchner.05a,Komijani.18,Cai.19}
and, more recently, in the physical $N=2$ case \cite{Cai.19} have shown such a singular charge 
dynamics. Intriguingly, this type of charge dynamical 
scaling in models of the Kondo limit smoothly connects to the $\omega/T$ scaling 
for the charge dynamics in the beyond-Landau-type quantum criticality in the mixed-valence regime \cite{Pixley.12}.

Epitaxial thin films of members of the 115 family and CeIn$_3$ have been available for some time \cite{Shishido.10} 
but STM measurements on these films of  CeCoIn$_5$ and CeRhIn$_5$ have 
been reported only very recently \cite{Haze.18,Haze.19}. 
The observed onset of the hybridization gap in the STM spectrum demonstrates the high-energy
$T_0$ scale which, as we have emphasized, is consistent with a Kondo destruction ground state in CeRhIn$_5$.
It will be interesting to see  whether
a lattice mismatch between substrate and thin film might be used as a substitute for pressure tuning and to establish the range of $\omega/T$ scaling both within the general phase diagram and with respect to the type of correlator, {\itshape i.e.}, single-particle excitations, two-particle correlators like the density-density, spin-spin, or current-current correlation functions, and their $n$-point ($n>4$) counterparts.

We also briefly discussed in Sec.\ \ref{sec:ARPES-STM} that the underlying assumption in the interpretation of STM spectra in terms of the equilibrium  local DOS is less justified at higher voltages. This may be particularly pertinent near the QCP, where the temperature of the measurement itself is expected to set the only relevant scale  \cite{Kirchner.09}. It would be interesting to explore the scaling of  spectral density with  bias voltage in the nonequilibrium regime which could be yet another way of unraveling the properties of the underlying QCP \cite{Ribeiro.15,Zamani.16b}.  

We have so far focused on YbRh$_2$Si$_2$ and Ce-115 compounds. It will be instructive to carry out 
measurements of 
   \qs{single-particle}
    properties in other candidate heavy-electron materials for 
Kondo destruction
\cite{QCNP13,Gegenwart.08,Si13.1,Ste01.1}. A case in point is CeNiAsO, a heavy-electron relative of the high-T$_c$ Fe-based oxypnictides.
Here  the recent neutron scattering experiments provide evidence for a local-moment antiferromagnetic order, whose ordering wave vector is determined by the RKKY interaction mediated by the conduction electron states near the small Fermi surface ({\itshape i.e.}, the Fermi surface of the conduction electrons alone, with the $4f$ electrons localized)
\cite{Wu-prl.19},
and transport measurements have suggested the possibility of a Kondo destruction QCP induced by either pressure or P-for-As doping \cite{Luo14.1}.

 More broadly, there is the question of where to look for new examples of
 Kondo destruction criticality. 
 If
  the $f-c$ hybridization is too strong, magnetic order  
  would more likely be
   of the SDW type that, when tuned to $T=0$, would result in a conventional QCP. 
Thus, weaker hybridization is expected to be a more favorable setting to access a possible Kondo destruction QCP.
Alternatively,  a low carrier density gives a small Fermi surface in a Kondo lattice and
 delays the full development of a Kondo singlet state with decreasing temperature. 
 CeNi$_2$As$_{2-\delta}$ appears to be an example of such a case
 with evidence of Kondo-destruction
 quantum criticality \cite{Luo.15}. 
 Finally, in the absence of tuning hybridization or carrier density, increasing frustration, 
 whether through crystal structure or reduced  dimensionality, 
 offers an exciting opportunity for discovering new examples \cite{Si.06,Fritsch.14,Tokiwa.15,Zhao.19}.

%
\section{Conclusion}

We have reviewed and compared recent ARPES and STM investigations on heavy-electron materials close to magnetic instabilities with a focus on 
Kondo destruction
quantum criticality. Real-space and momentum-space spectroscopies combine
the power of both methods \cite{Crepaldi.13,Nicoara.06}   which has proven to be useful in the study of complex materials such as the
cuprate
high-temperature superconductors \cite{Markiewicz.04,Shen.08} and the Kondo insulator SmB$_6$ \cite{Matt.18}.
In the context of 
cerium- and ytterbium-based
rare earth intermetallics as well as actinide-based compounds, such a combination seems particularly promising given that much of the excitement and interest generated by these materials derives from the interplay of local and itinerant degrees of freedom. While Kondo screening is primarily a local phenomenon, a possible Fermi-volume increase is best addressed in momentum space. Method-specific constraints, limited energy resolution and  the need for very low temperatures in order to resolve a Fermi momentum change across a 
Kondo destruction
 quantum-critical point pose unique challenges to both ARPES and STM investigations.
 
On the other hand, combining ARPES and STM results with other  measurements, like resistivity and magnetotransport measurements, neutron scattering and optical conductivity investigations, can provide a consistent picture of 
Kondo destruction
quantum criticality that emerges as a function of some nonthermal tuning parameter and  enables one to locate a specific compound in the general phase diagram of heavy-electron materials. 
This appears particularly relevant in the present context in order to aide a separation of bulk and surface contributions as both ARPES and STM are primarily surface sensitive. The change in symmetry and $c-f$ hybridization that typically occurs at surfaces can in Kondo systems substantially modify low-energy scales as compared to their bulk value.

We have emphasized the distinction between the spectroscopic properties that reflect the high-energy Kondo physics, 
such as the formation of the hybridization gap, and those that are capable of probing the nature of quantum criticality, such as  low-temperature 
isothermal measurements across the quantum-critical point. The latter has become possible in the STM measurements of YbRh$_2$Si$_2$, which corroborates the 
Kondo destruction
energy scale that had been extracted by isothermal magnetotransport and thermodynamic measurements. In CeRhIn$_5$, strong evidence for Kondo destruction in the one-electron excitation spectrum
has been provided by quantum oscillation measurements across the critical pressure.
It will certainly be instructive to explore further signatures of beyond-Landau quantum criticality in these and other heavy-electron systems.

\subsection*{Acknowledgements} 
We are grateful 
to the late Elihu Abrahams,
 Jim Allen, Pegor Aynajian, 
 Ang Cai,
 Kai Grube, Nigel Hussey, Kevin Ingersent,
 Johann Kroha, Hilbert von L\"ohneysen, 
 Yuji Matsuda,
 Emilian Nica, Jed H. Pixley, Pedro Ribeiro, Rong Yu, Zuo-Dong Yu, \mbox{Farzaneh Zamani}, and Gertrud Zwicknagl for useful discussions. 
 Part of these discussions took place at the 2018 Hangzhou Workshop on Quantum Matter and we thank all participants
 of the workshop.
This work was in part supported by 
the National Key R\&D Program of the MOST of China, Grant No.\ 2016YFA0300200 (S.K., Q.C., and D.F.), 
the National Science Foundation of China, No.\ 11774307 (S.K.), and No.\ 11874330 (Q.C.), and the Science Challenge Project,  Grant  No. TZ2016004 (D.F.).
\sk{The work in Vienna was supported by the FWF (Project No. P 29296-N27) and the European Microkelvin Platform (H2020 Project No. 824109).}
The work at Los Alamos was performed under the auspices of the U.S. Department of Energy, Division of Materials Sciences and Engineering.
The work at Rice was in part supported
by the NSF (DMR-1920740) and the Robert A. Welch Foundation (C-1411).
S.K.\ acknowledges support by MOST of Taiwan, Grant No. 108-2811-M-009-500 and hospitality of NCTU, \sk{Hsinchu}.
Donglai Feng is supported by the Anhui Initiative in Quantum Information Technologies.
Q.S.\ acknowledges 
the hospitality  and the support by a Ulam Scholarship 
from the Center for Nonlinear Studies at Los Alamos National Laboratory and the hospitality of the Aspen
Center for Physics (NSF, PHY-1607611).

%


\end{document}